\documentclass[prd,aps,floats,preprintnumbers,preprint]{revtex4}
\usepackage{graphicx}
\usepackage[english]{babel} 
\usepackage{braket}
\newcommand{\be}{\begin{equation}}
\newcommand{\ee}{\end{equation}}
\newcommand{\bea}{\begin{eqnarray}}
\newcommand{\eea}{\end{eqnarray}}

\usepackage{morefloats} 
\usepackage[titletoc,title]{appendix}
\usepackage{color}
\begin{document}

\title{Effects of local  features of the inflaton potential on the spectrum and bispectrum of primordial perturbations}

\author{Alexander Gallego Cadavid$^{1,3}$, Antonio Enea Romano$^{1,2,3}$, Stefano Gariazzo${}^{4,5}$}
\affiliation
{
${}^{1}$Instituto de Fisica, Universidad de Antioquia, A.A.1226, Medellin, Colombia\\
${}^{2}$Department of Physics, University of Crete, 71003 Heraklion,Greece \\
${}^{3}$Yukawa Institute for Theoretical Physics, Kyoto University, Japan\\
${}^{4}$Department of Physics, University of Torino, Via P. Giuria 1, I--10125 Torino, Italy\\
${}^{5}$INFN, Sezione di Torino, Via P. Giuria 1, I--10125 Torino, Italy
}
\begin{abstract}
We study the effects of a class of features of the potential of slow-roll inflationary models corresponding to a step symmetrically dumped by an even power negative exponential factor, which we call local features.
Local-type features differ from other branch-type features considered previously, because the potential is only affected in a limited range of the scalar field value, and are symmetric respect to the location of the feature.  
This type of features only affects the spectrum and bispectrum in a narrow range of scales which leave the horizon during the time interval corresponding to the modification of the potential. On the contrary branch-type features have effects on all the perturbation modes leaving the horizon when the field value is within the interval defining the branch, introducing for example differences in the power spectrum between large and small scale which are absent in the case of local-type features.

The spectrum and bispectrum of primordial curvature perturbations are affected by oscillations around the scale $k_0$ exiting the horizon at the time $\tau_0$ corresponding to the feature.
We also compute the effects of the features on the CMB temperature and polarization spectra, showing the effects or different choices of parameters.
\end{abstract}

\maketitle

\section{Introduction}
Theoretical cosmology has entered in the last decades in a new era in which different models can be compared directly to high-precision observations \cite{et, wmapcpr, pxvi,Ade:2015xua}. One fundamental source of information about the early Universe is the cosmic microwave background (CMB) radiation, which according to the standard cosmological model consists of the photons that decoupled from the primordial plasma when the neutral hydrogen atoms started to form.  

According to the inflation theory \cite{anewtype,encyclopaedia} the CMB temperature anisotropies arose from primordial curvature perturbation, whose 
spectrum is approximately scale invariant.
An approximately scale invariant spectrum of curvature perturbation, with a small tilt, provides a good fit of CMB data \cite{Ade:2015xua},
but recent analyses of the WMAP and Planck data 
have shown evidence of a feature around the scale $k=0.002$ Mpc${}^{-1}$ in the power spectrum of primordial scalar fluctuations 
\cite{Shafieloo:2003gf,Nicholson:2009pi,Hazra:2013ugu,Hazra:2014jwa,
Nicholson:2009zj,Hunt:2013bha,Hunt:2015iua,
Goswami:2013uja,Matsumiya:2001xj,Matsumiya:2002tx,
Kogo:2003yb,Kogo:2005qi,Nagata:2008tk,Ade:2015lrj,
Gariazzo:2014dla,DiValentino:2015zta,Hazra:2014aea},
that correspond to a dip in the CMB temperature spectrum at $l\simeq20$.
This kind of feature of the curvature perturbations spectrum provides an important observational motivation to find theoretical models able to explain it.
In  particular in this paper we will consider the effects of features of the inflaton potential in single field inflationary model as possible explanation of the features of the power spectrum. 

The effects of features of the inflaton potential were first studied by Starobinsky \cite{starobinsky}, and  CMB  data have shown  some  glitches of 
the power spectrum \cite{constraints1,constraints2} compatible with these features 
\cite{Starobinsky:1998mj,Joy:2007na,Joy:2008qd,Mortonson:2009qv,Novaes:2015uza}.

The Starobinsky model and its generalizations \cite{Bousso:2013uia,Romano:2014kla,GallegoCadavid:2015bsn,whipped,wiggly} belong to a class of branch 
features (BF) which involve a step function or a smoothed version of the latter \cite{Adams}, and consequently introduce a distinction between a left 
and right branch of the potential.
In this paper instead we will consider the effects of local features (LF) \cite{GallegoThesis} which only modify the potential locally in  field 
space, while leaving it unaffected sufficiently far from the feature. The important consequence is that also the effects of LF on the spectrum and 
bispectrum are local, while BF modify the spectrum in a wider range of scales. Features of the inflaton potential could be produced by different 
sources \cite{Palma:2014hra,Mooij:2015cxa}, for example particle production \cite{pp}, or phase transitions \cite{Adams2}, but here we will only 
study their effects adopting a phenomenological approach, as done originally by Starobinsky in his seminal work.
\section{Single field slow-roll inflation}
\begin{figure}
 \begin{minipage}{.45\textwidth}
  \includegraphics[scale=0.6]{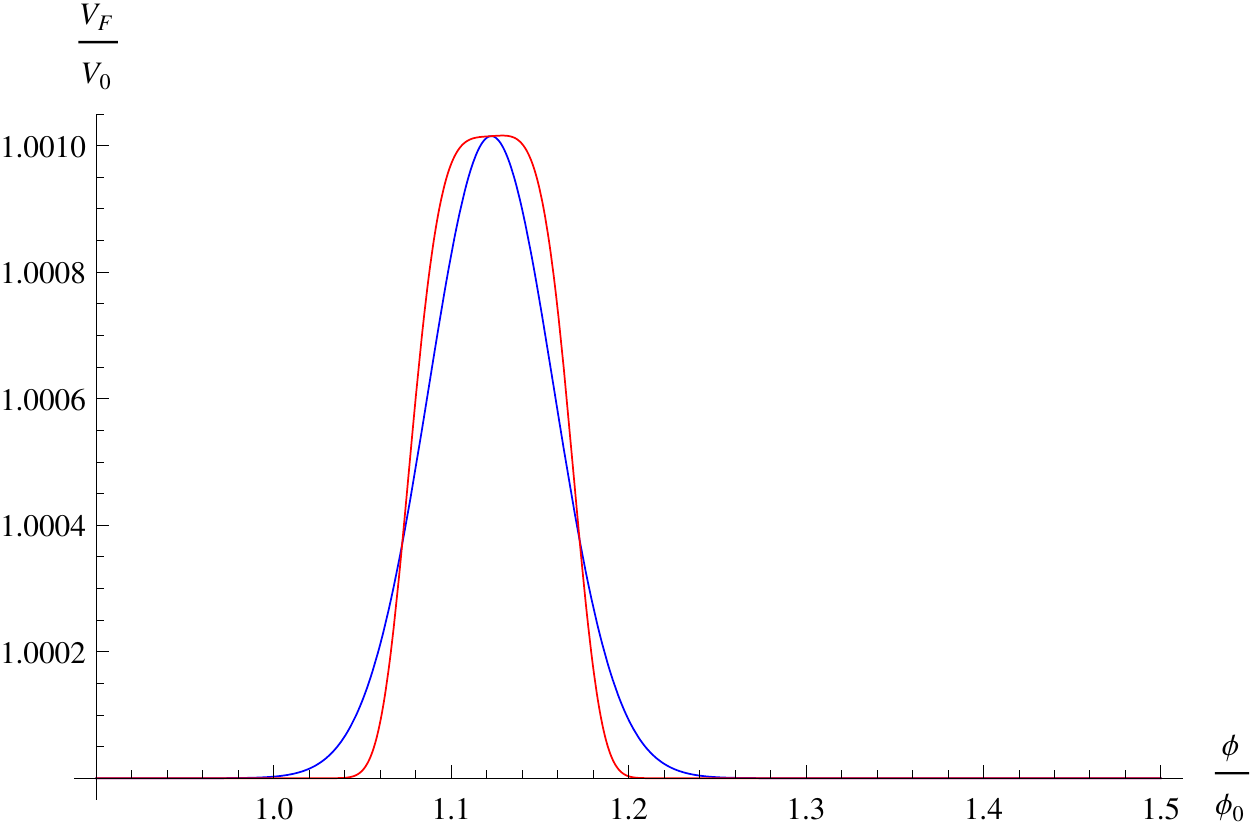}
  \end{minipage}
 \begin{minipage}{.45\textwidth}
  \includegraphics[scale=0.6]{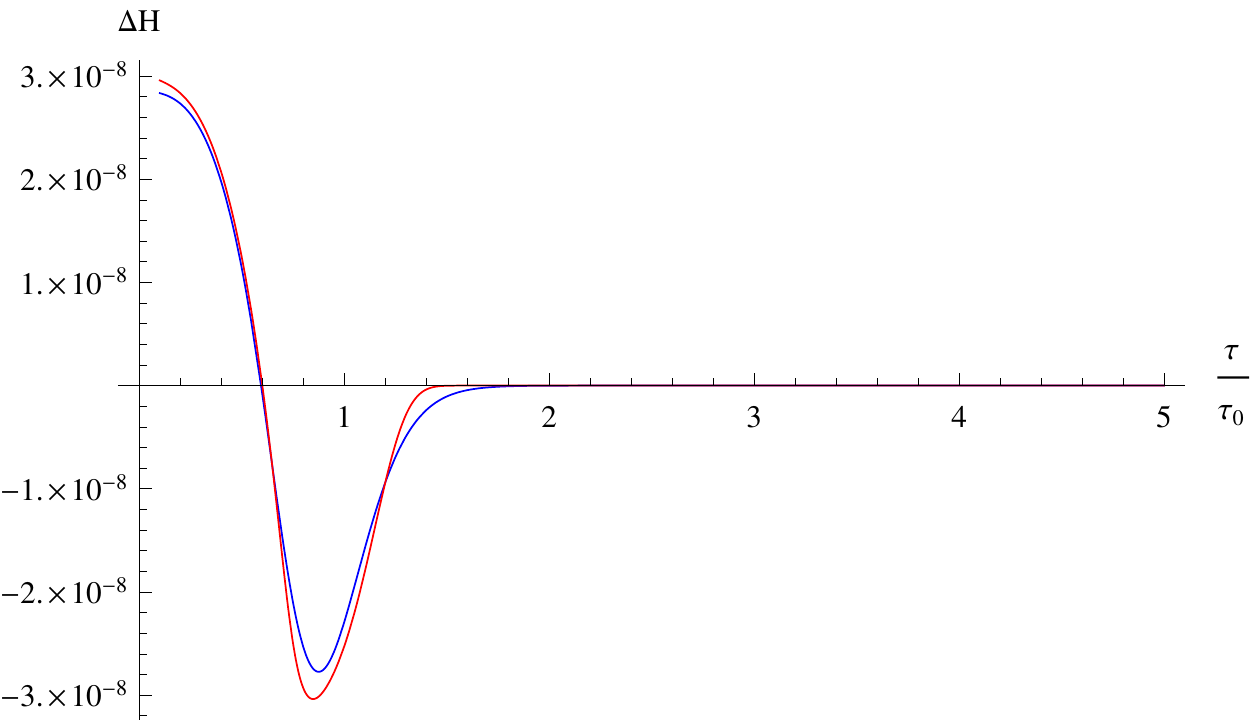}
 \end{minipage}
 \begin{minipage}{.45\textwidth}
  \includegraphics[scale=0.6]{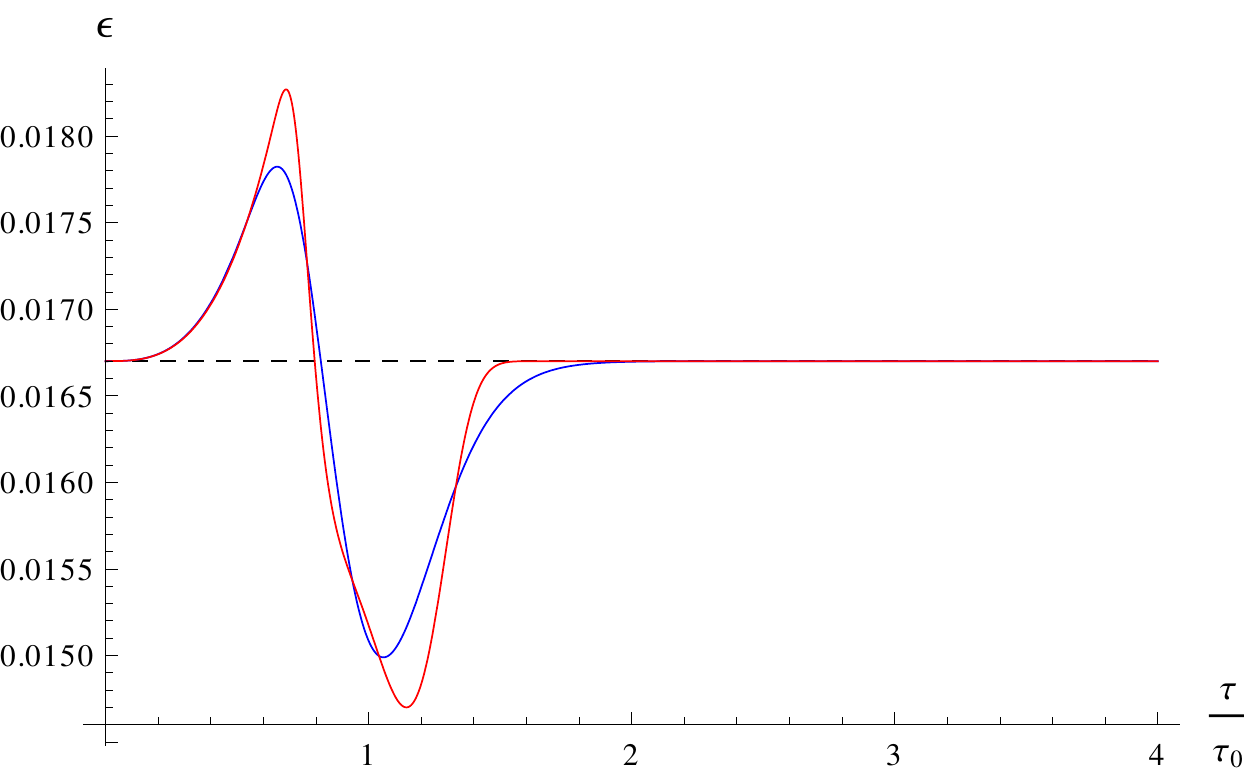}
  \end{minipage}
 \begin{minipage}{.45\textwidth}
  \includegraphics[scale=0.6]{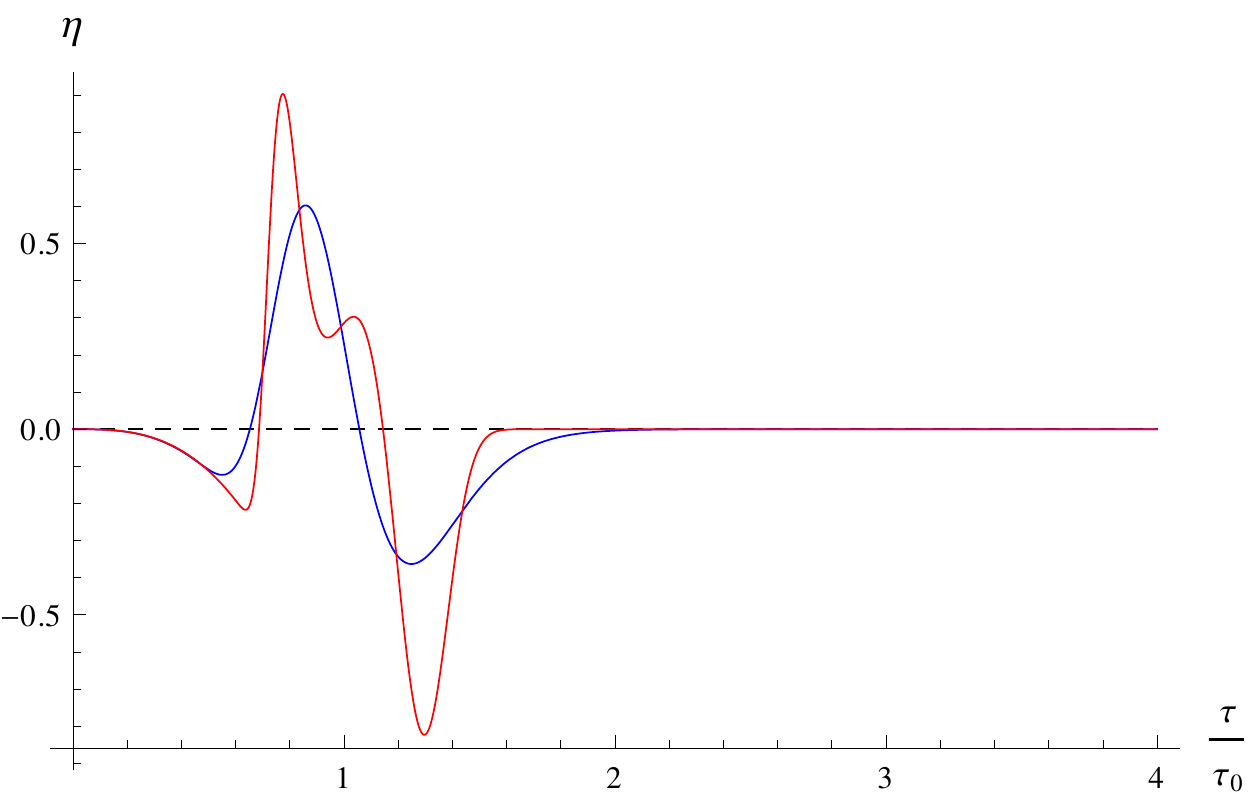}
 \end{minipage}
 \caption{From left to right and top to bottom the numerically computed $V_F/V_0, \Delta H, \epsilon$ and $\eta$ are plotted for $\lambda=10^{-11}$, $\sigma=0.05$  and $n=1$ (blue) and $n=2$ (red). The dashed black lines correspond to the featureless behavior.}
\label{nback}
\end{figure}

\begin{figure}
 \begin{minipage}{.45\textwidth}
  \includegraphics[scale=0.6]{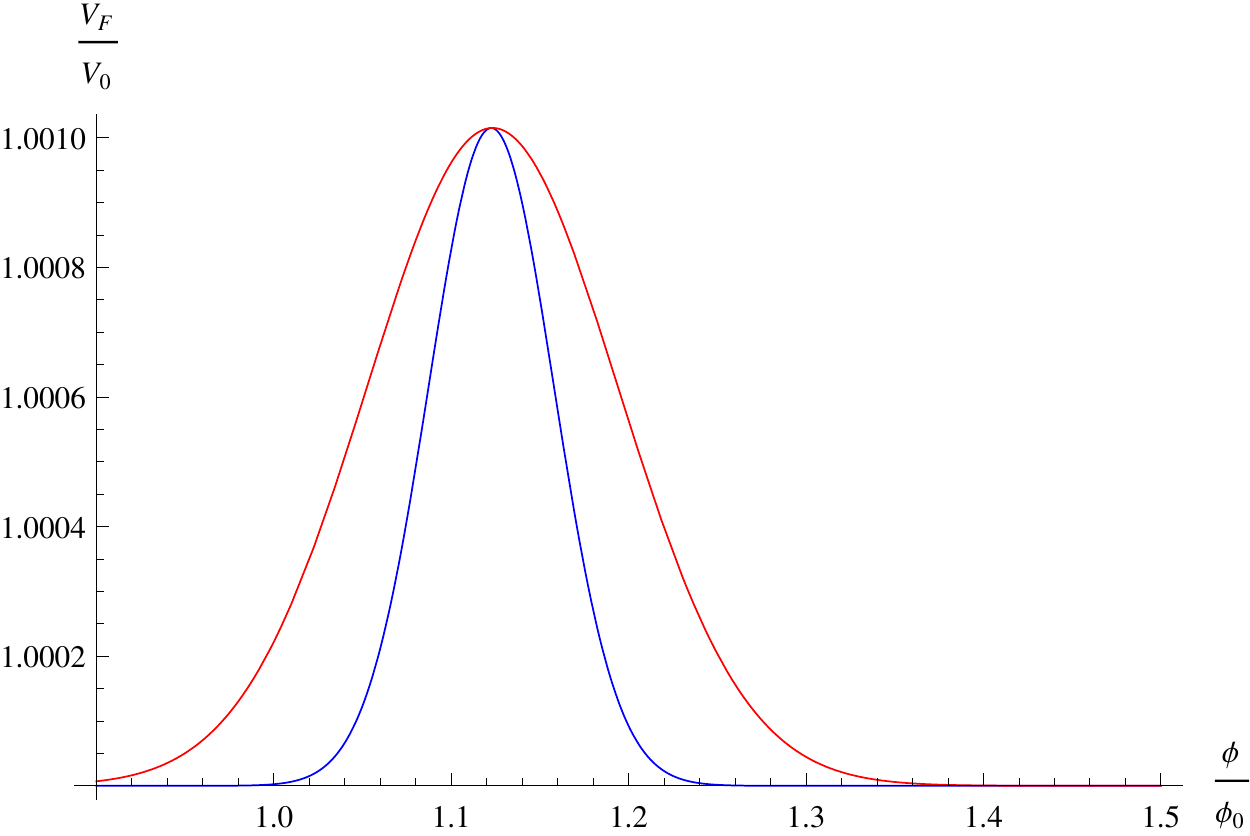}
  \end{minipage}
 \begin{minipage}{.45\textwidth}
  \includegraphics[scale=0.6]{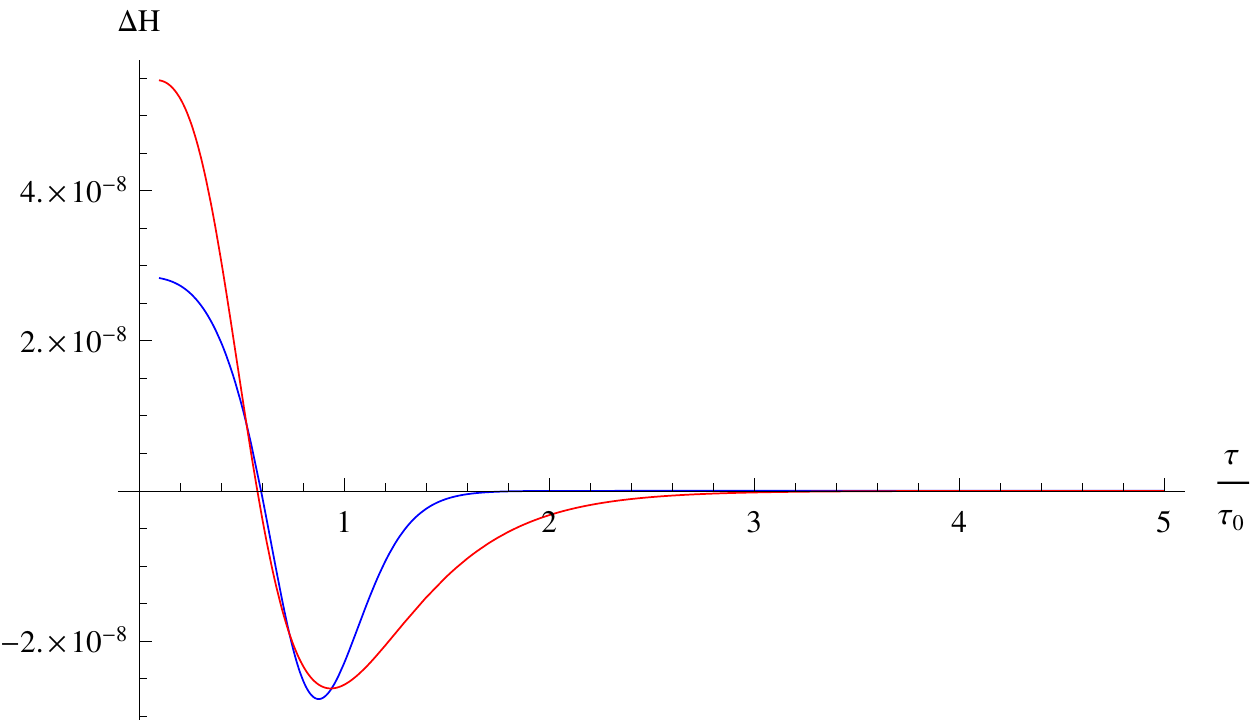}
 \end{minipage}
 \begin{minipage}{.45\textwidth}
  \includegraphics[scale=0.6]{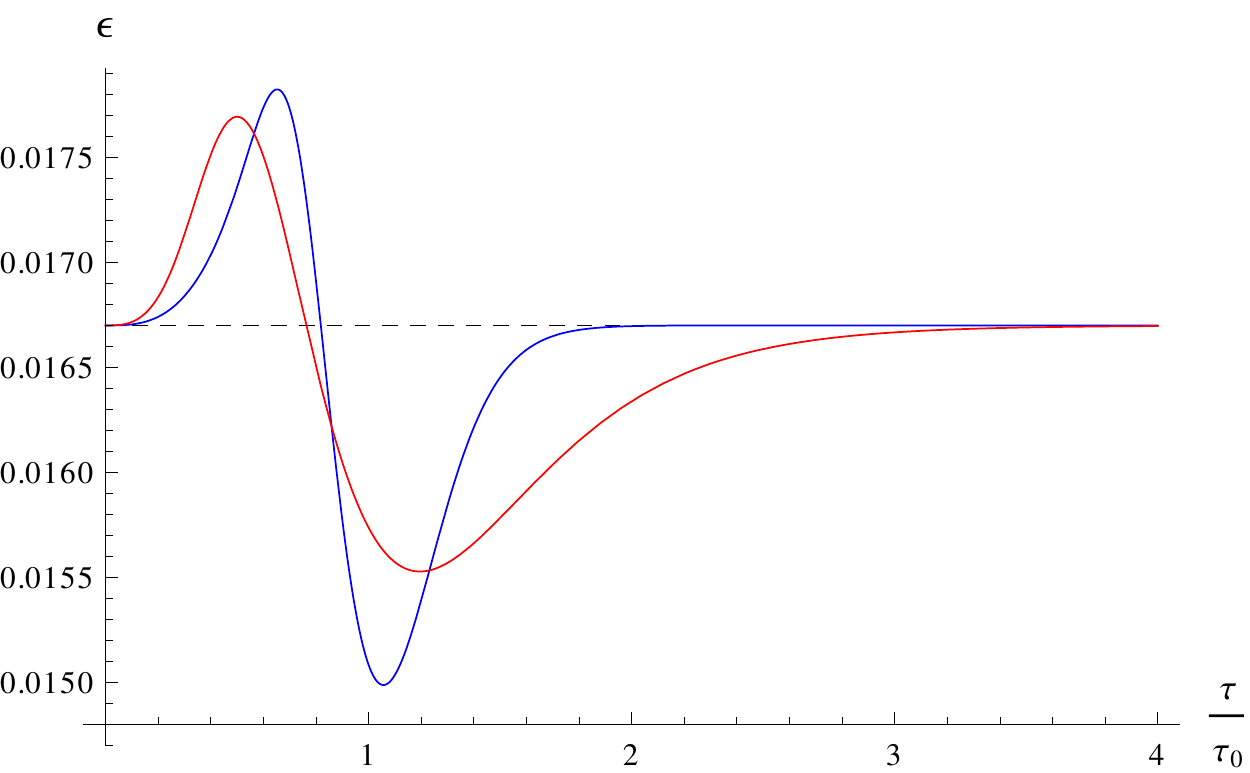}
  \end{minipage}
 \begin{minipage}{.45\textwidth}
  \includegraphics[scale=0.6]{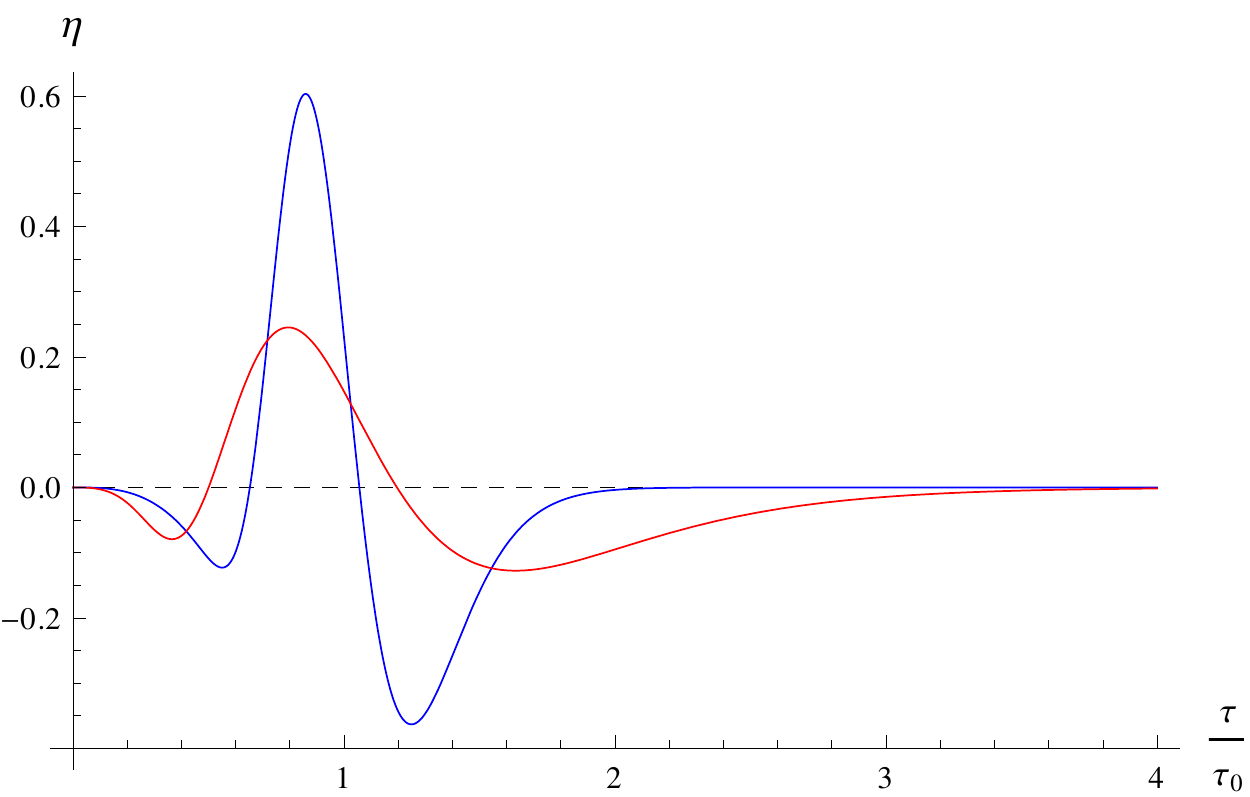}
 \end{minipage}
 \caption{From left to right and top to bottom the numerically computed $V_F/V_0, \Delta H, \epsilon$ and $\eta$ are plotted for $\lambda=10^{-11}$, $\sigma=0.05$ (blue), $\sigma=0.1$ (red) and $n=1$. The dashed black lines correspond to the featureless behavior.}
\label{sigmaback}
\end{figure}

\begin{figure}
 \begin{minipage}{.45\textwidth}
  \includegraphics[scale=0.6]{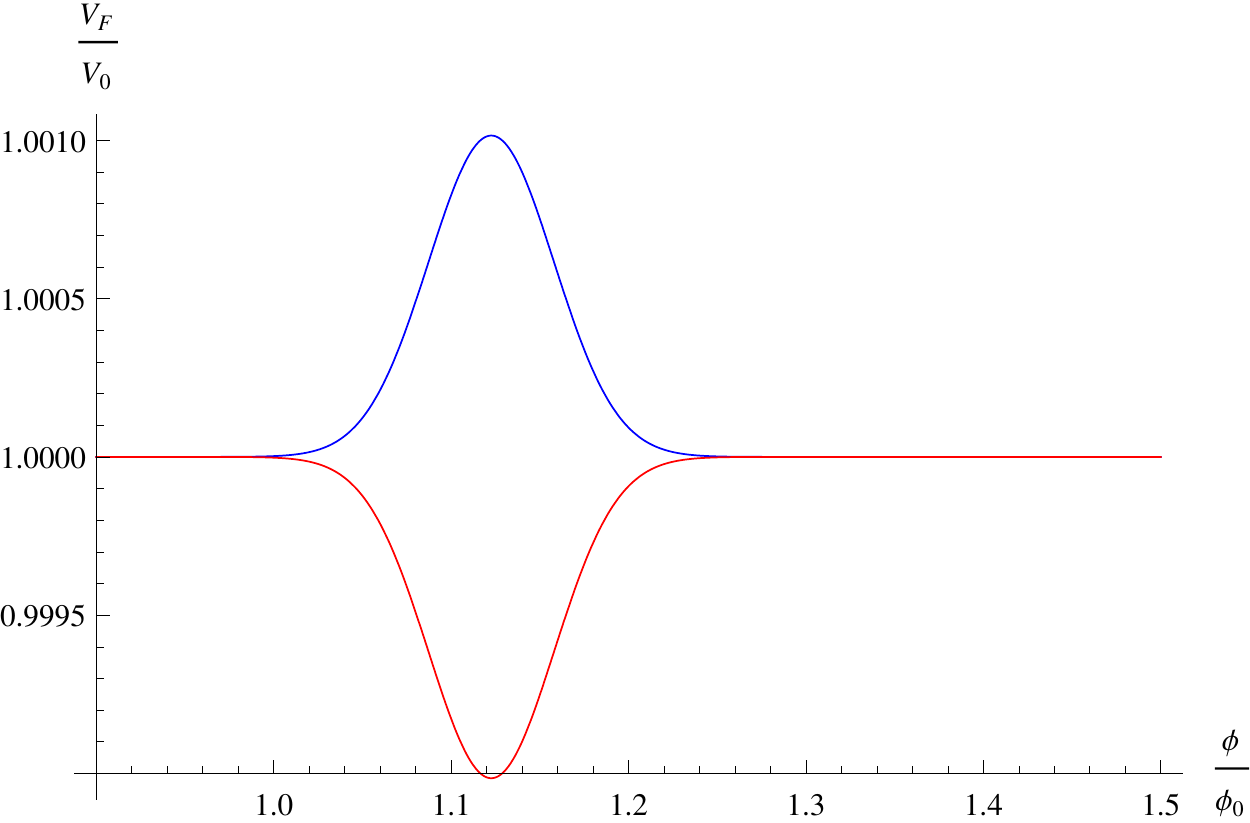}
  \end{minipage}
 \begin{minipage}{.45\textwidth}
  \includegraphics[scale=0.6]{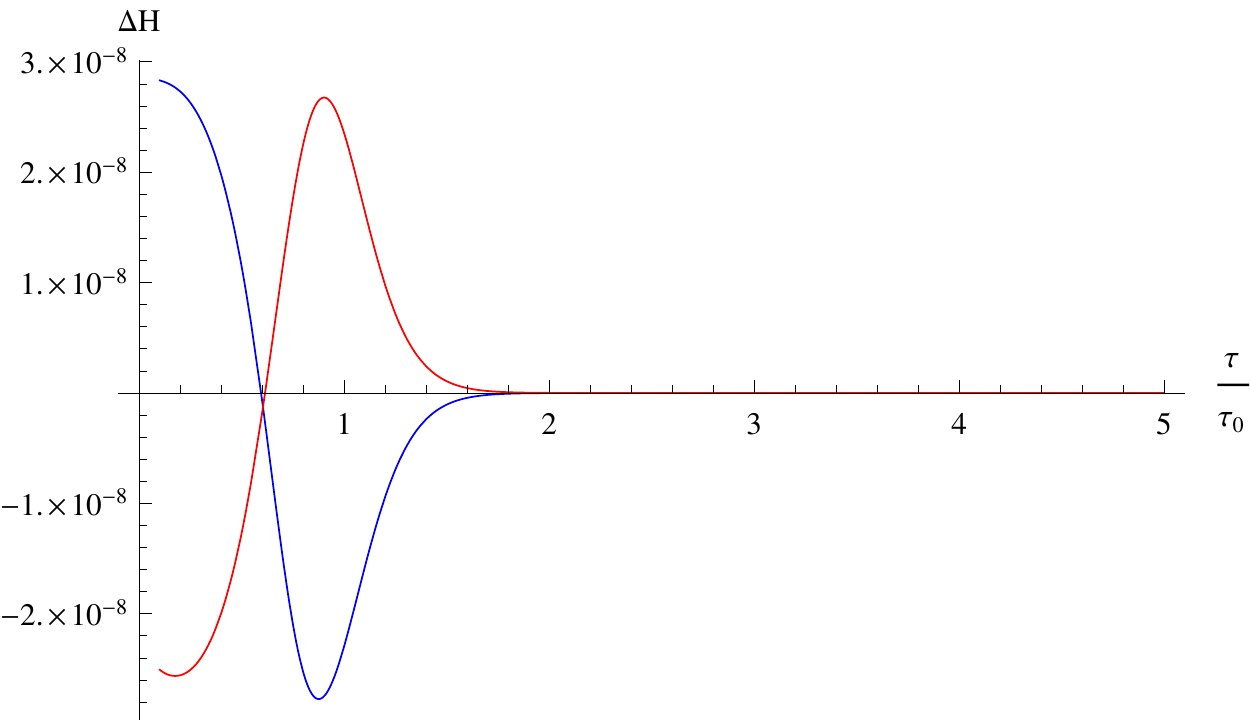}
 \end{minipage}
 \begin{minipage}{.45\textwidth}
  \includegraphics[scale=0.6]{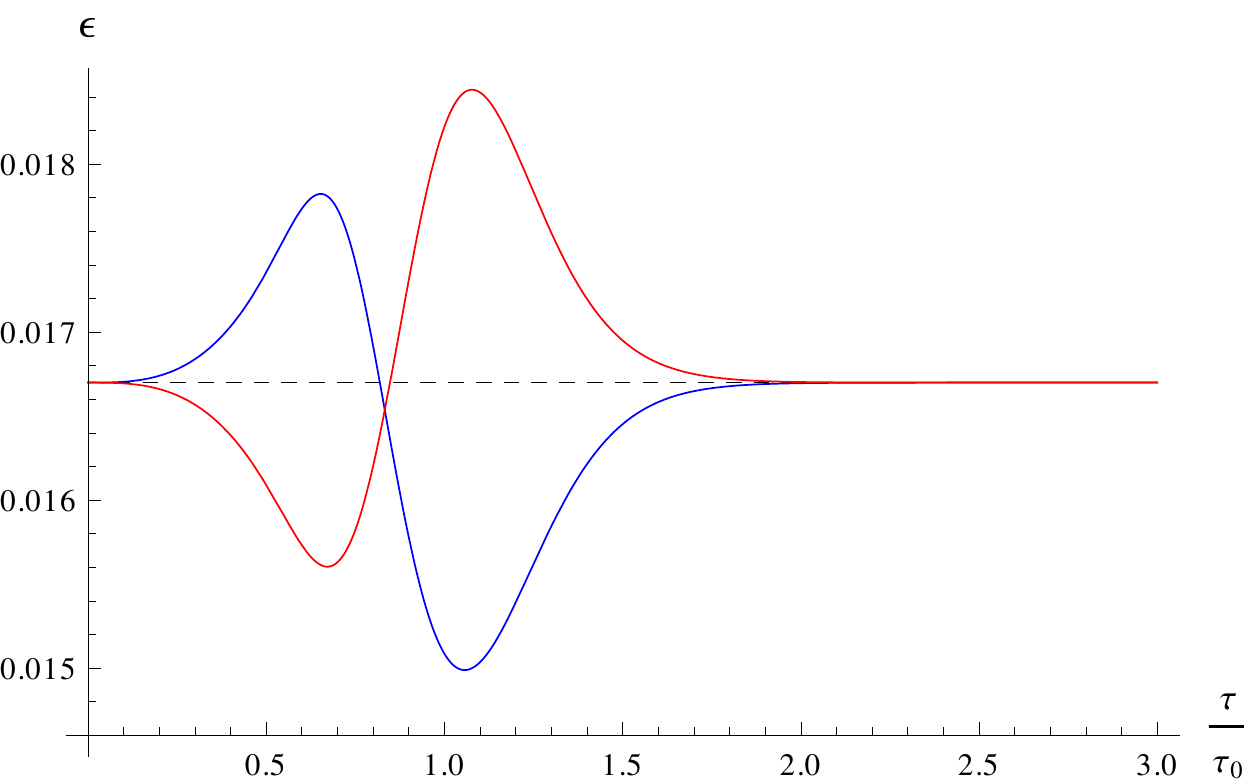}
  \end{minipage}
 \begin{minipage}{.45\textwidth}
  \includegraphics[scale=0.6]{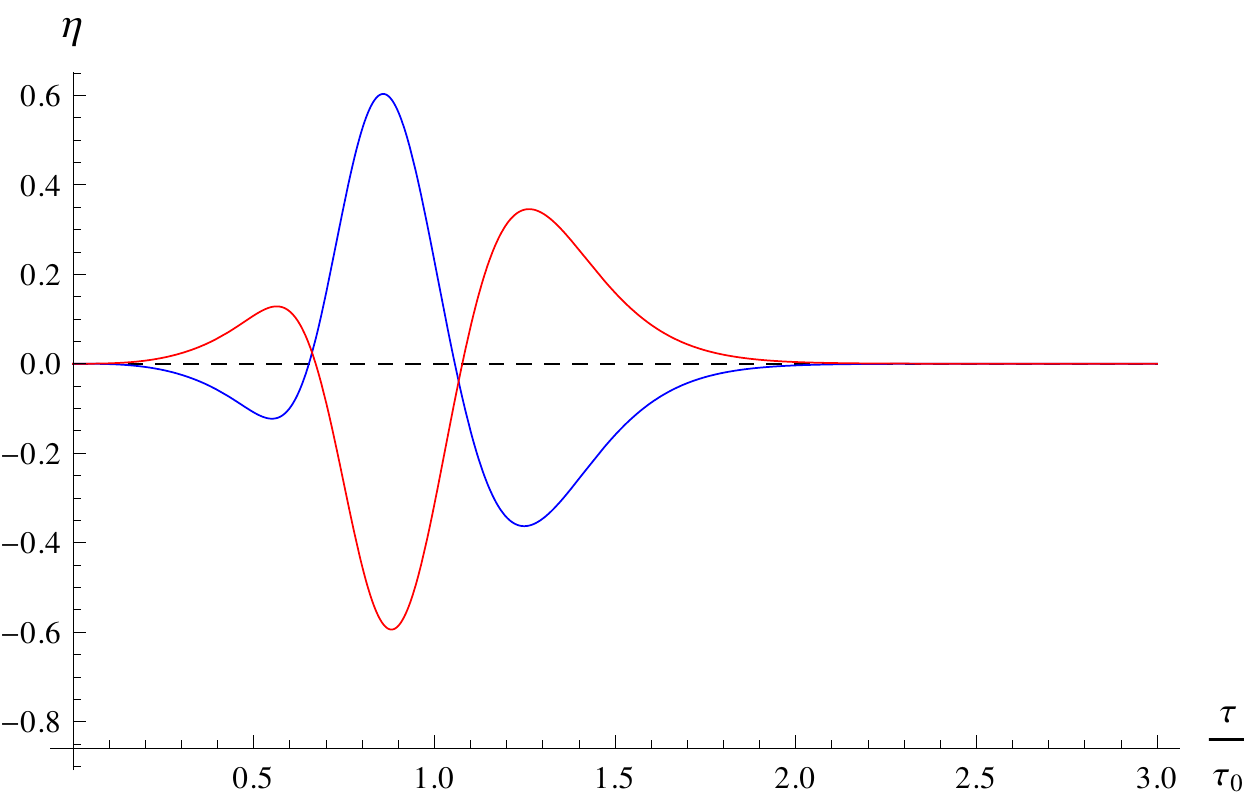}
 \end{minipage}
 \caption{From left to right and top to bottom the numerically computed $V_F/V_0, \Delta H, \epsilon$ and $\eta$ are plotted for $\lambda=10^{-11}$ (blue) and $-\lambda=10^{-11}$ (red), $\sigma=0.05$ and $n=1$. The dashed black lines correspond to the featureless behavior.}
\label{pmlambdaback}
\end{figure}

\begin{figure}
 \begin{minipage}{.45\textwidth}
  \includegraphics[scale=0.6]{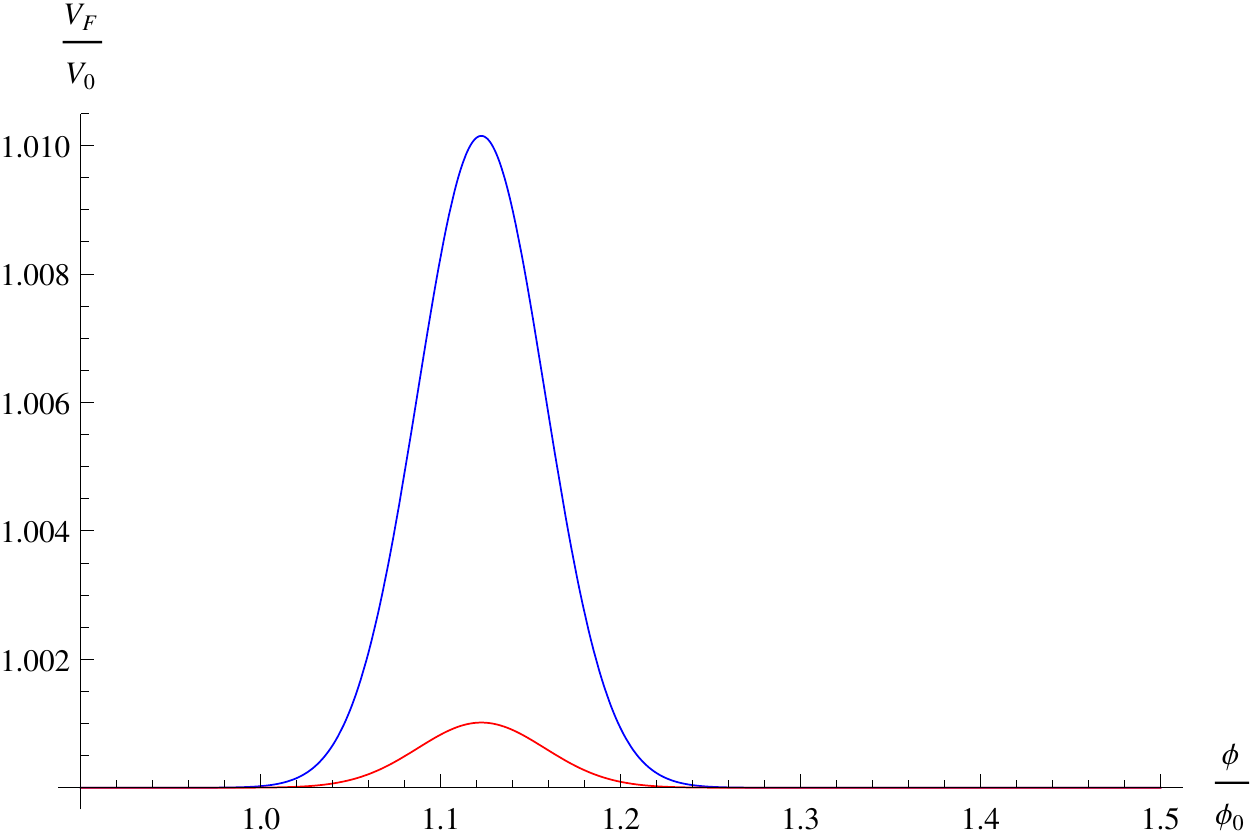}
  \end{minipage}
 \begin{minipage}{.45\textwidth}
  \includegraphics[scale=0.6]{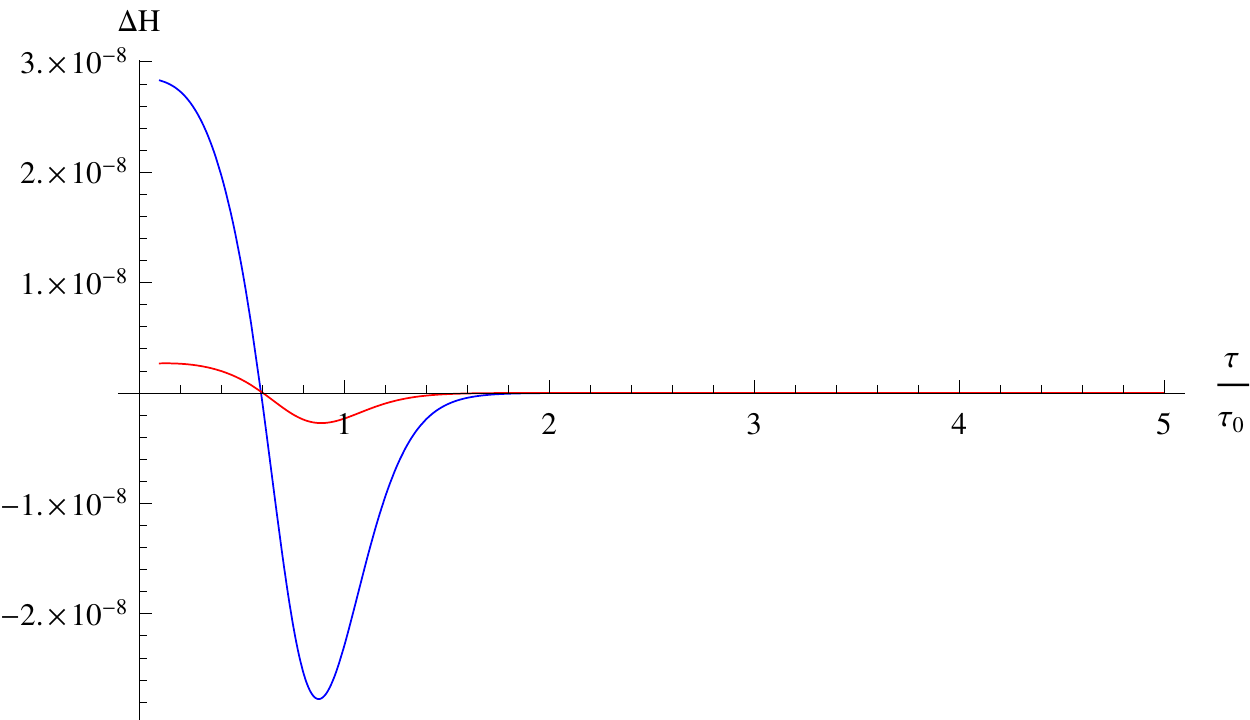}
 \end{minipage}
 \begin{minipage}{.45\textwidth}
  \includegraphics[scale=0.6]{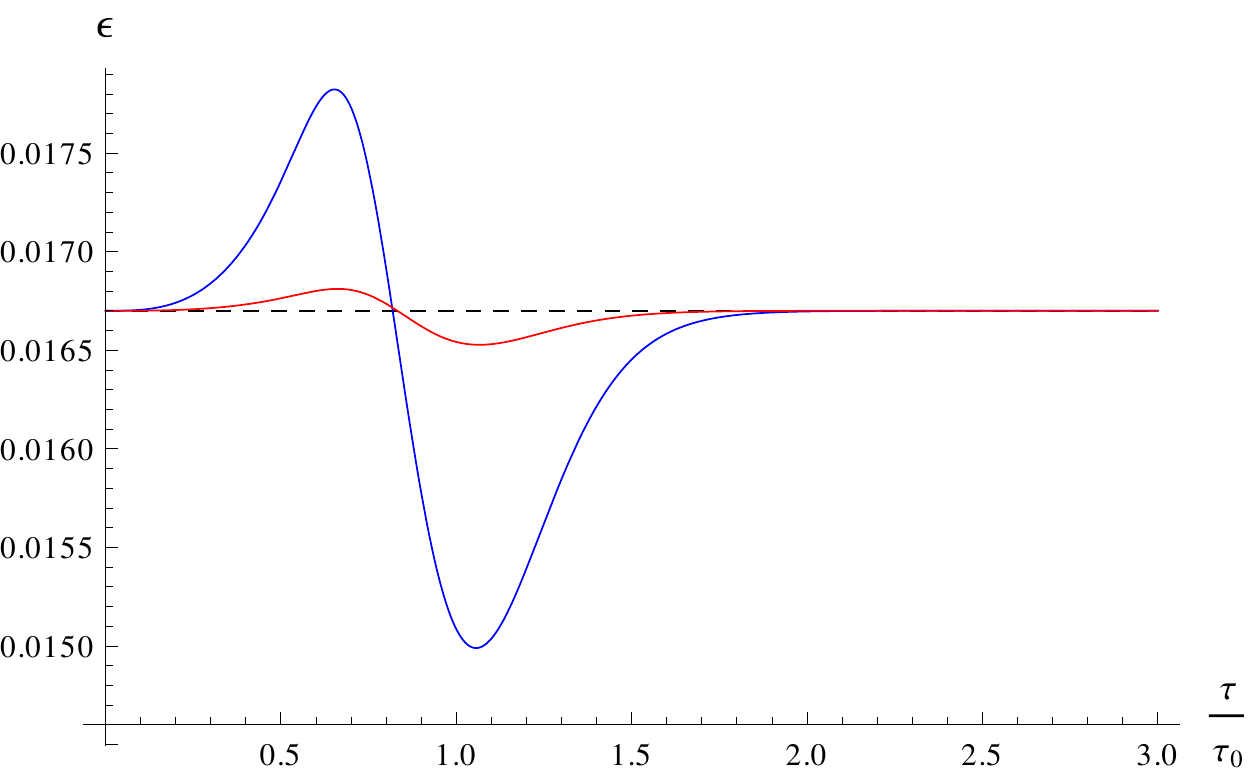}
  \end{minipage}
 \begin{minipage}{.45\textwidth}
  \includegraphics[scale=0.6]{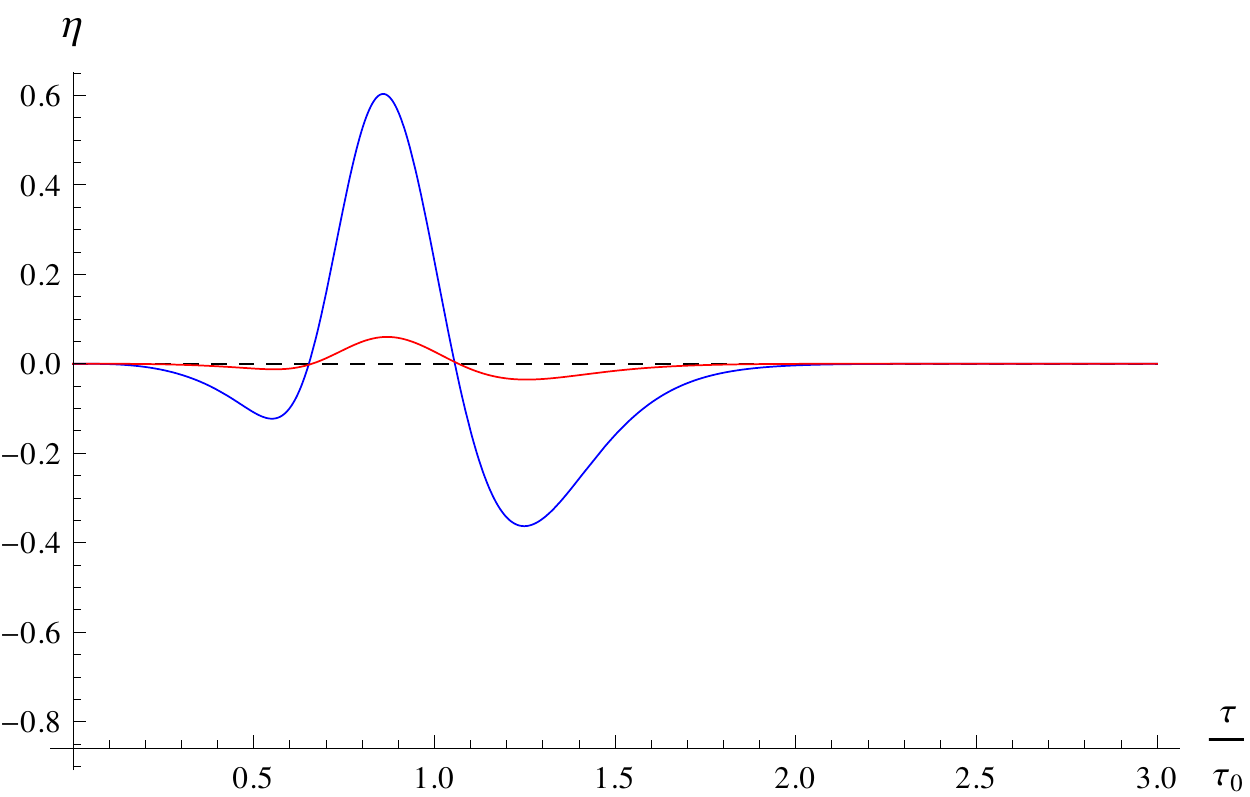}
 \end{minipage}
 \caption{From left to right and top to bottom the numerically computed $V_F/V_0, \Delta H, \epsilon$ and $\eta$ are plotted for $\lambda=10^{-11}$ (blue) and $\lambda=10^{-12}$ (red), $\sigma=0.05$ and $n=1$. The dashed black lines correspond to the featureless behavior.}
\label{plambdaback}
\end{figure}


\begin{figure}
 \begin{minipage}{.45\textwidth}
  \includegraphics[scale=0.6]{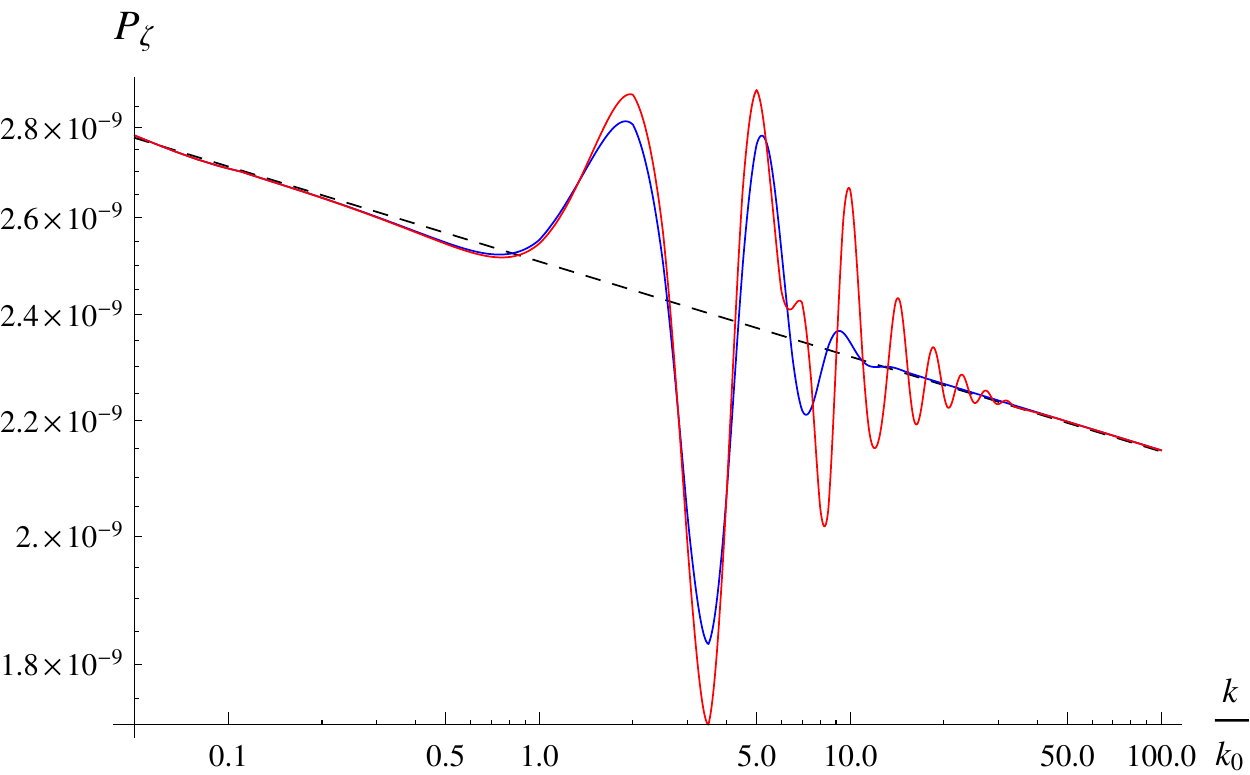}
  \end{minipage}
 \begin{minipage}{.45\textwidth}
  \includegraphics[scale=0.6]{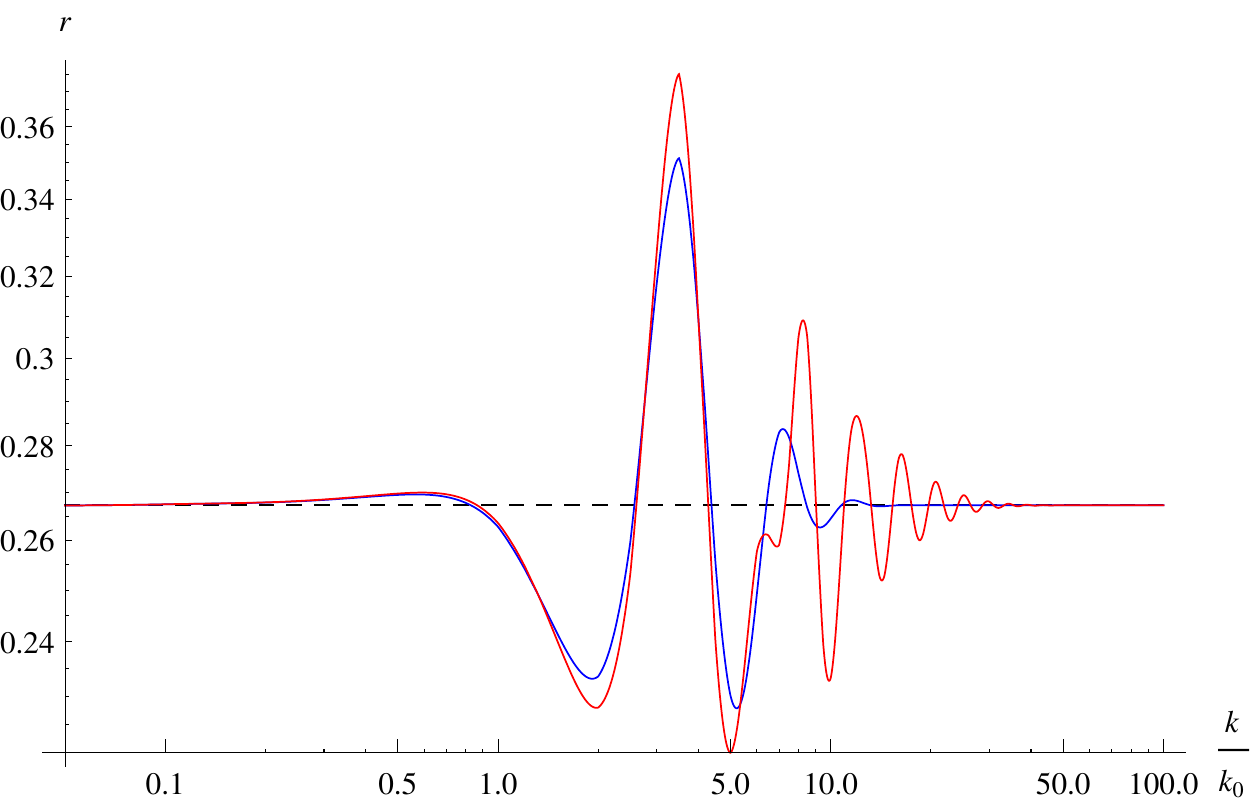}
 \end{minipage}
 \caption{The numerically computed $P_{\zeta}$ and $r$ are  plotted for $\lambda=10^{-11}$, $\sigma=0.05$  and $n=1$ (blue) and $n=2$ (red). The dashed black lines correspond to the featureless behavior.}
\label{npert}
\end{figure}

\begin{figure}
 \begin{minipage}{.45\textwidth}
  \includegraphics[scale=0.6]{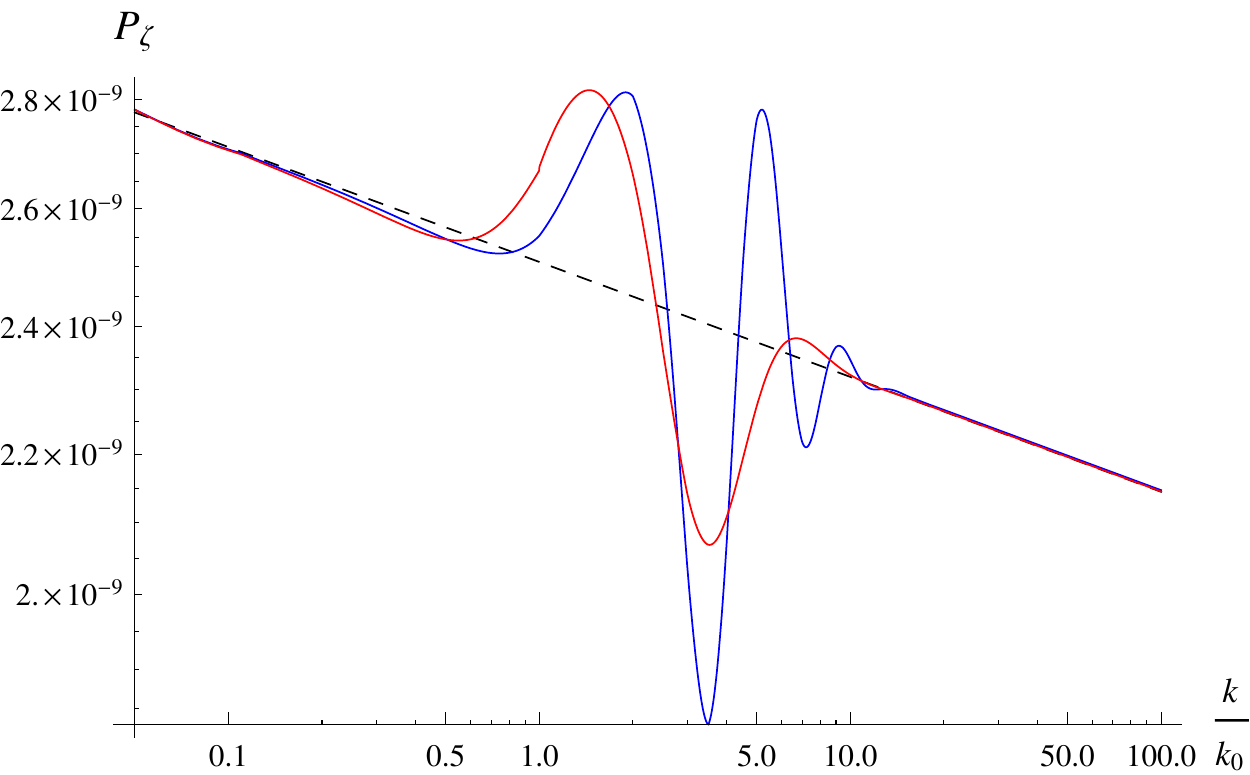}
  \end{minipage}
 \begin{minipage}{.45\textwidth}
  \includegraphics[scale=0.6]{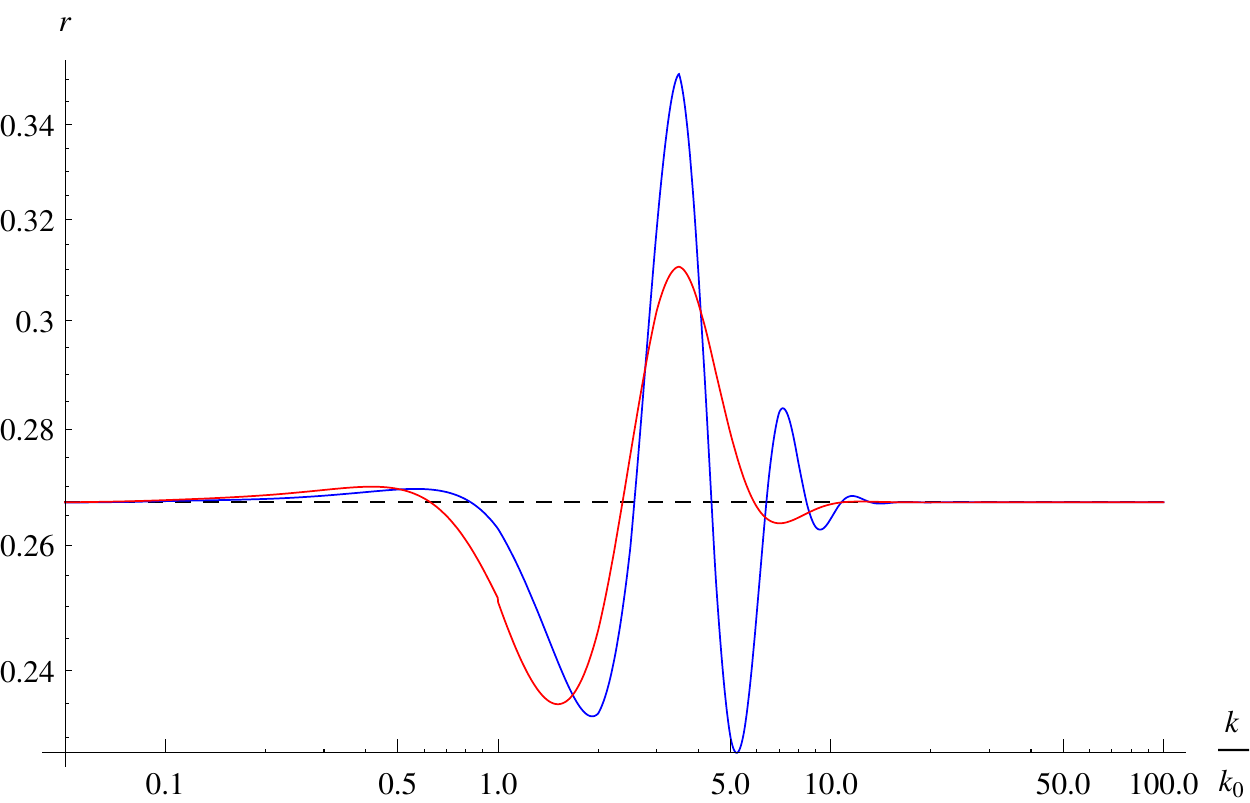}
 \end{minipage}
 \caption{From left to right the numerically computed $P_{\zeta}$ and $r$ are plotted for $\lambda=10^{-11}$, $\sigma=0.05$ (blue) $\sigma=0.1$ (red) and $n=1$. The dashed black lines correspond to the featureless behavior.}
\label{sigmapert}
\end{figure}

\begin{figure}
 \begin{minipage}{.45\textwidth}
  \includegraphics[scale=0.6]{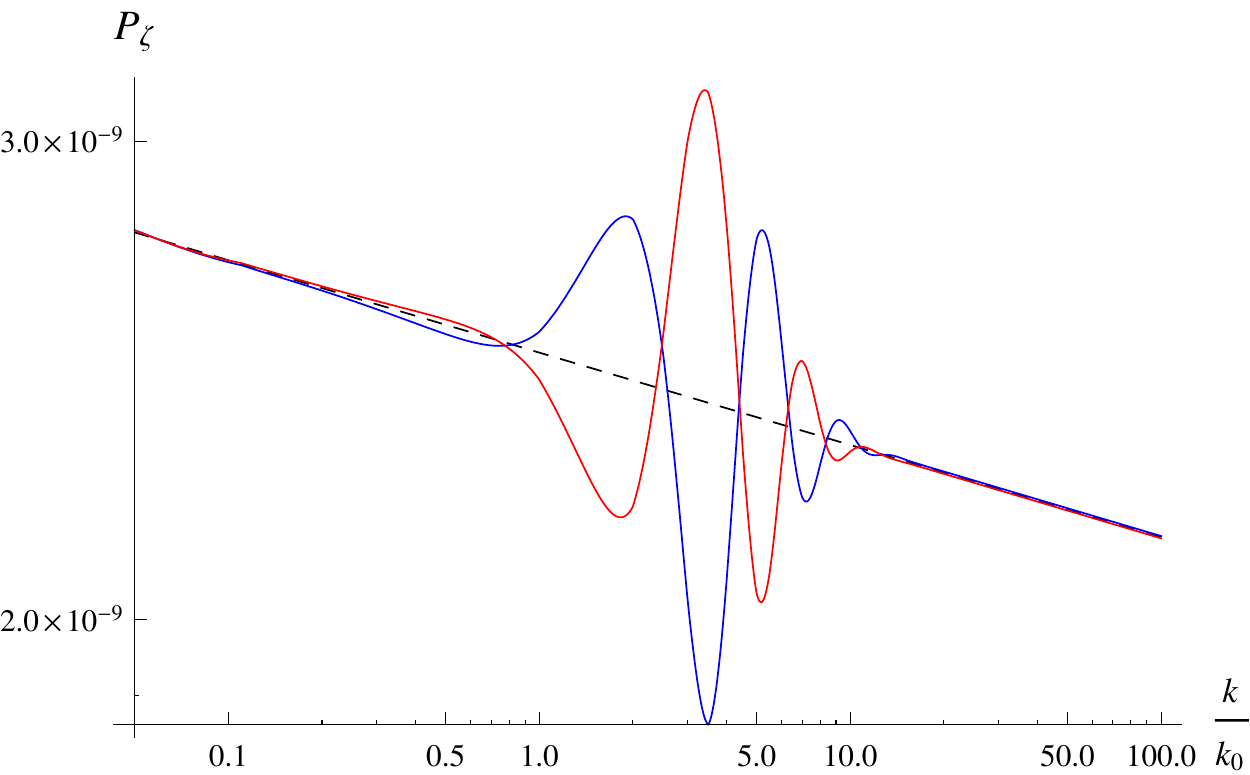}
  \end{minipage}
 \begin{minipage}{.45\textwidth}
  \includegraphics[scale=0.6]{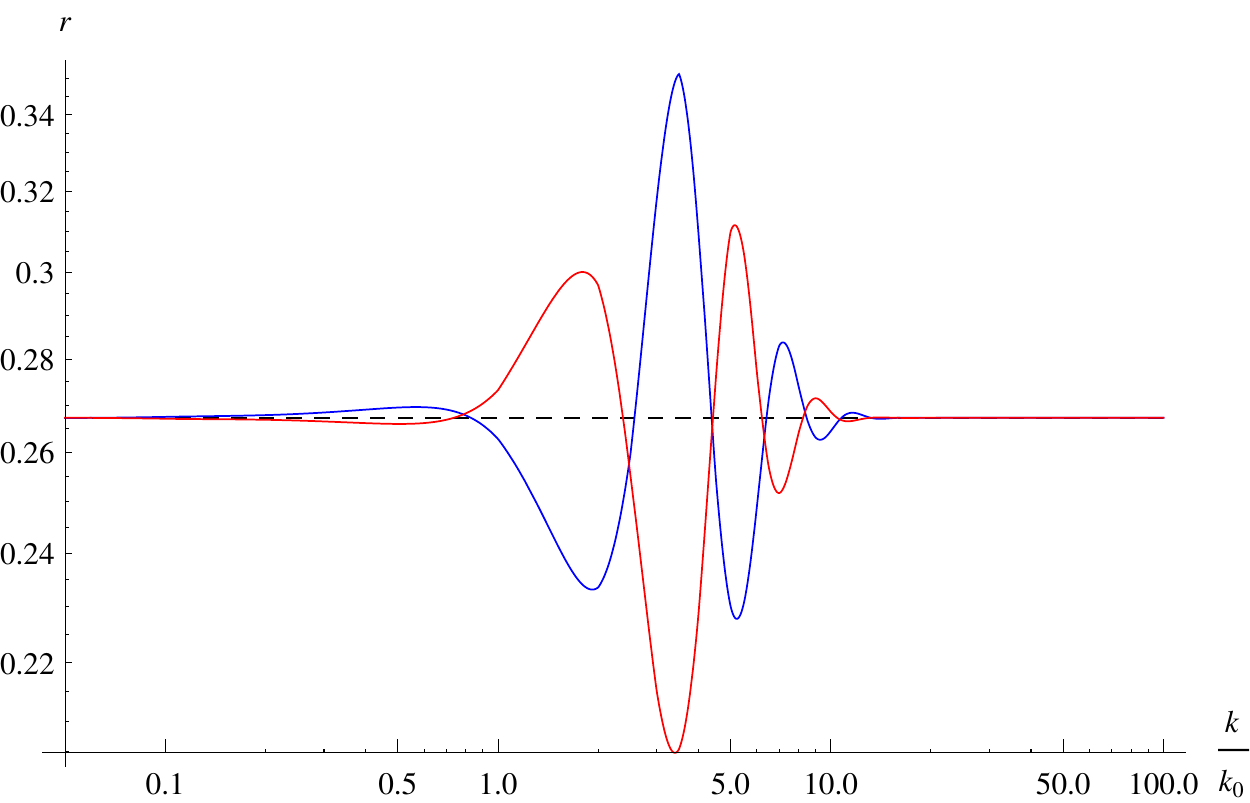}
 \end{minipage}
 \caption{The numerically computed $P_{\zeta}$ and $r$  are plotted for $\lambda=10^{-11}$ (blue) and $-\lambda=10^{-11}$ (red), $\sigma=0.05$ and $n=1$. The dashed black lines correspond to the featureless behavior.}
\label{pmlambdapert}
\end{figure}

\begin{figure}
 \begin{minipage}{.45\textwidth}
  \includegraphics[scale=0.6]{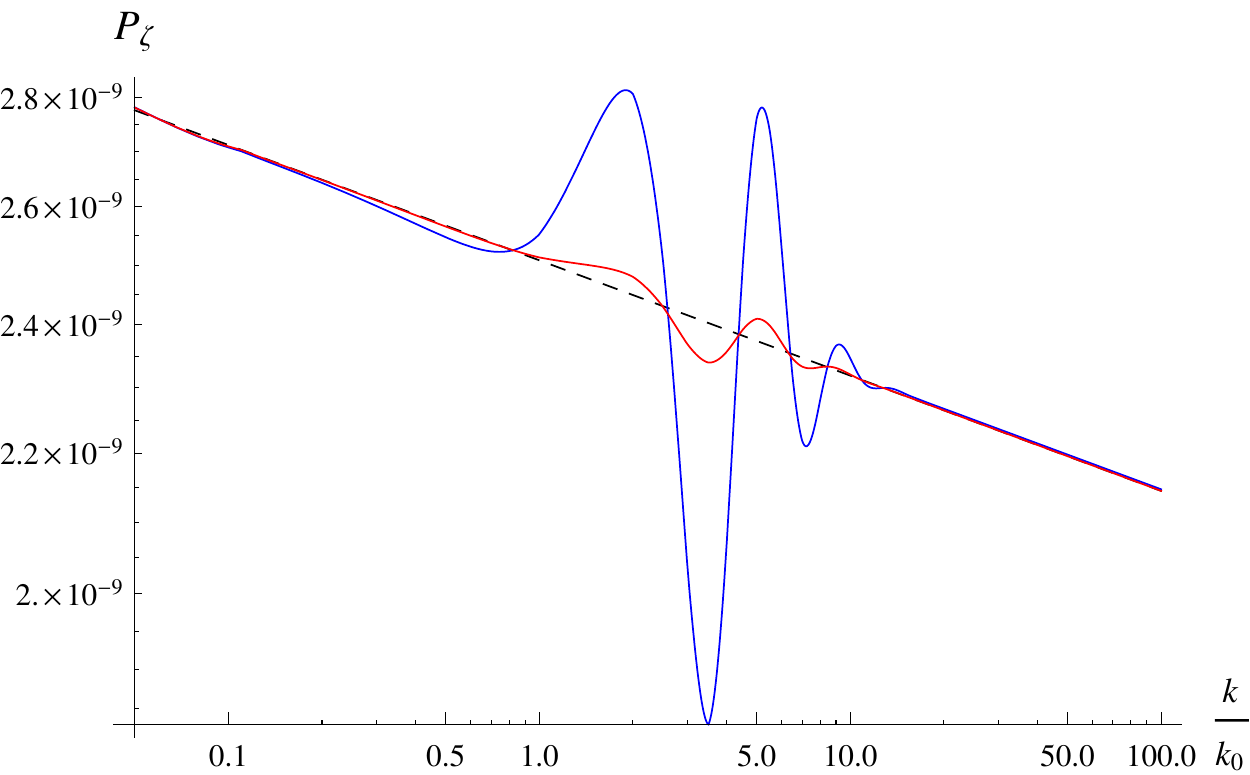}
  \end{minipage}
 \begin{minipage}{.45\textwidth}
  \includegraphics[scale=0.6]{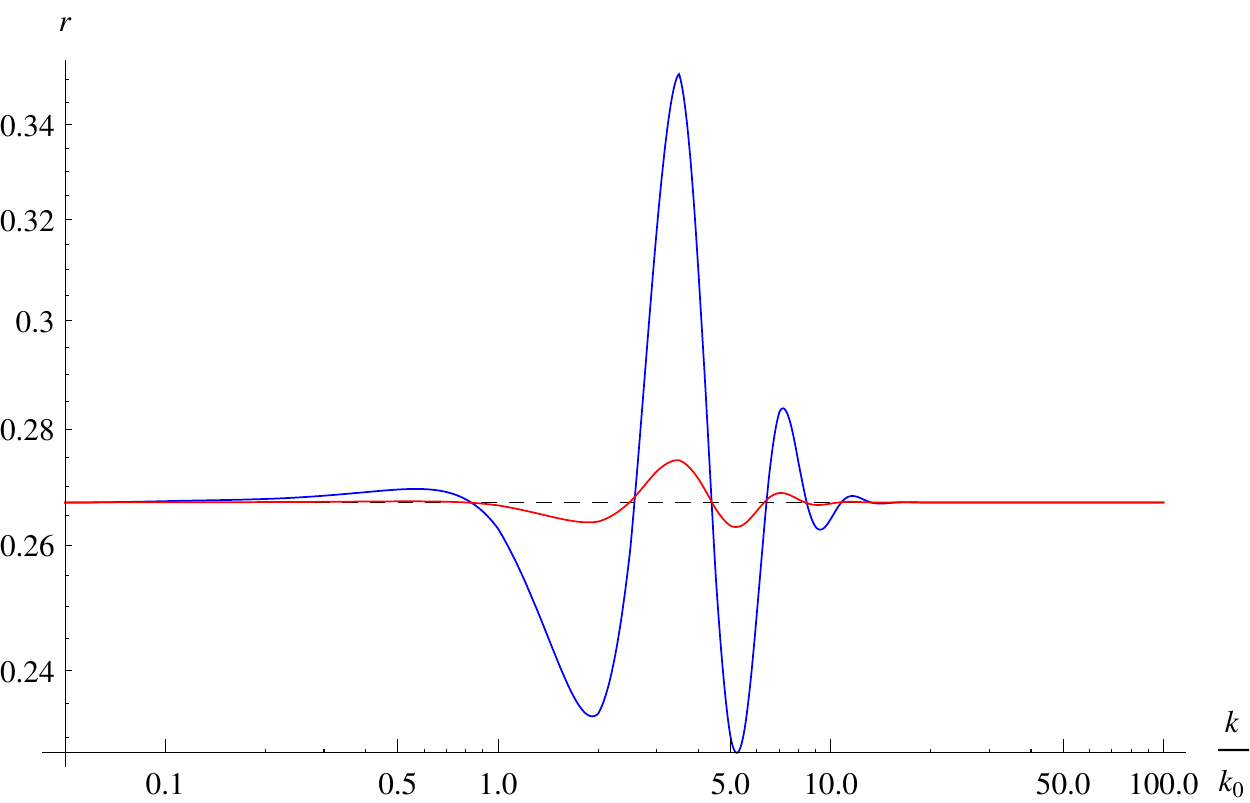}
 \end{minipage}
 \caption{The numerically computed $P_{\zeta}$ and $r$ are plotted for $\lambda=10^{-11}$ (blue) and $\lambda=10^{-12}$ (red), $\sigma=0.05$ and $n=1$. The dashed black lines correspond to the featureless behavior.}
\label{plambdapert}
\end{figure}

\begin{figure}
  \includegraphics[scale=1]{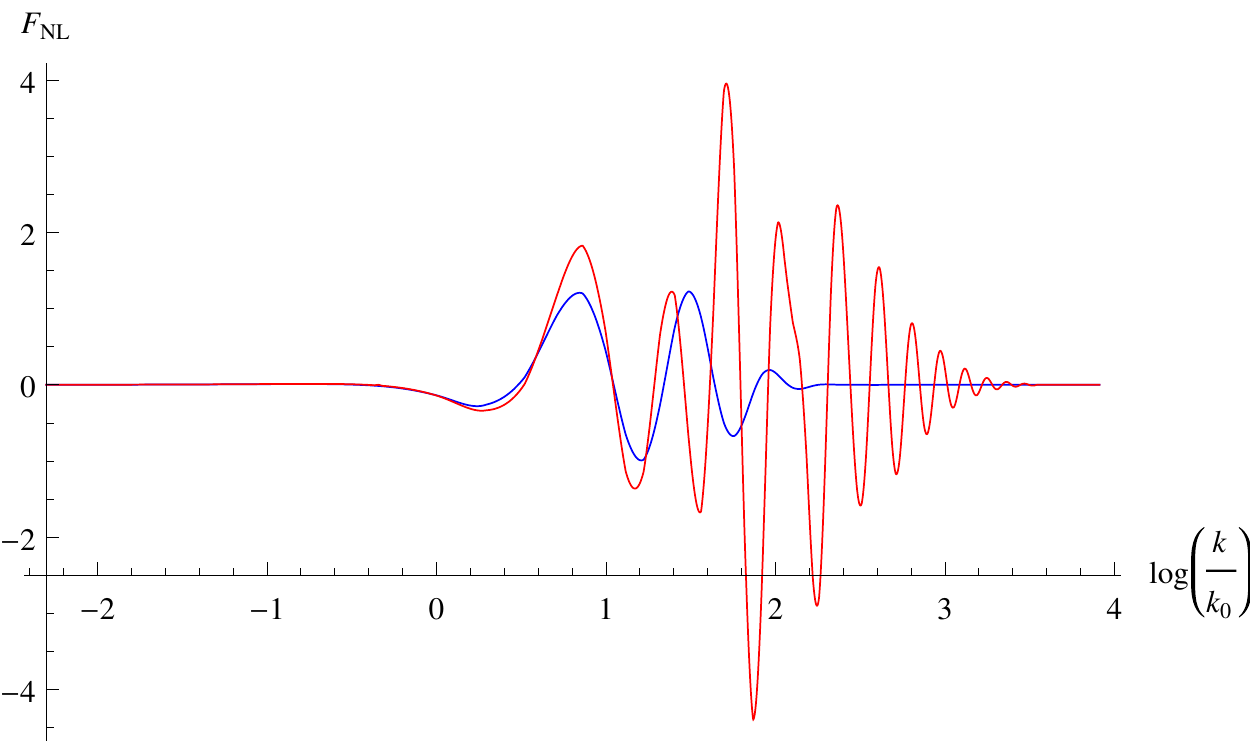}
 \caption{The numerically computed equilateral shape bispectrum is plotted for
 for $\lambda=10^{-11}$, $\sigma=0.05$  and $n=1$ (blue) and $n=2$ (red).}
\label{nequilateral}
\end{figure}

\begin{figure}
  \includegraphics[scale=1]{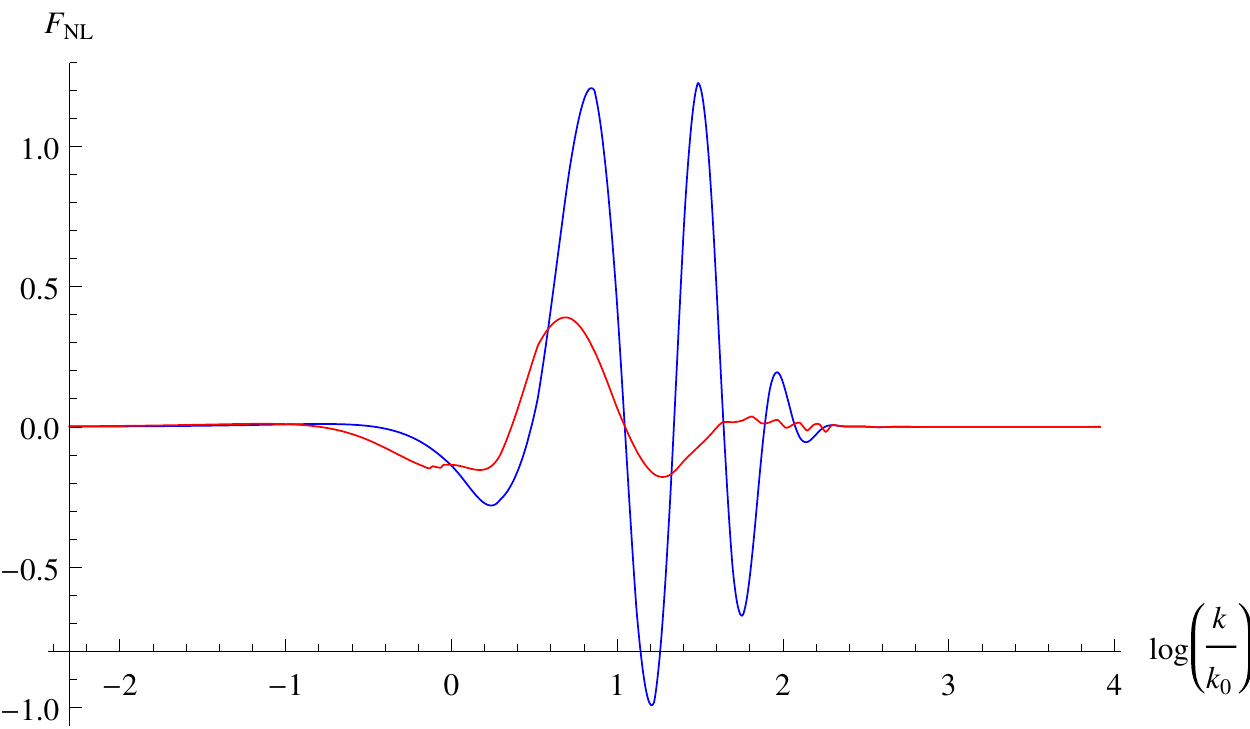}
 \caption{The numerically computed equilateral shape bispectrum is plotted for for $\lambda=10^{-11}$, $\sigma=0.05$ (blue) $\sigma=0.1$ (red) and $n=1$.}
\label{sigmaequilateral}
\end{figure}

\begin{figure}
  \includegraphics[scale=1]{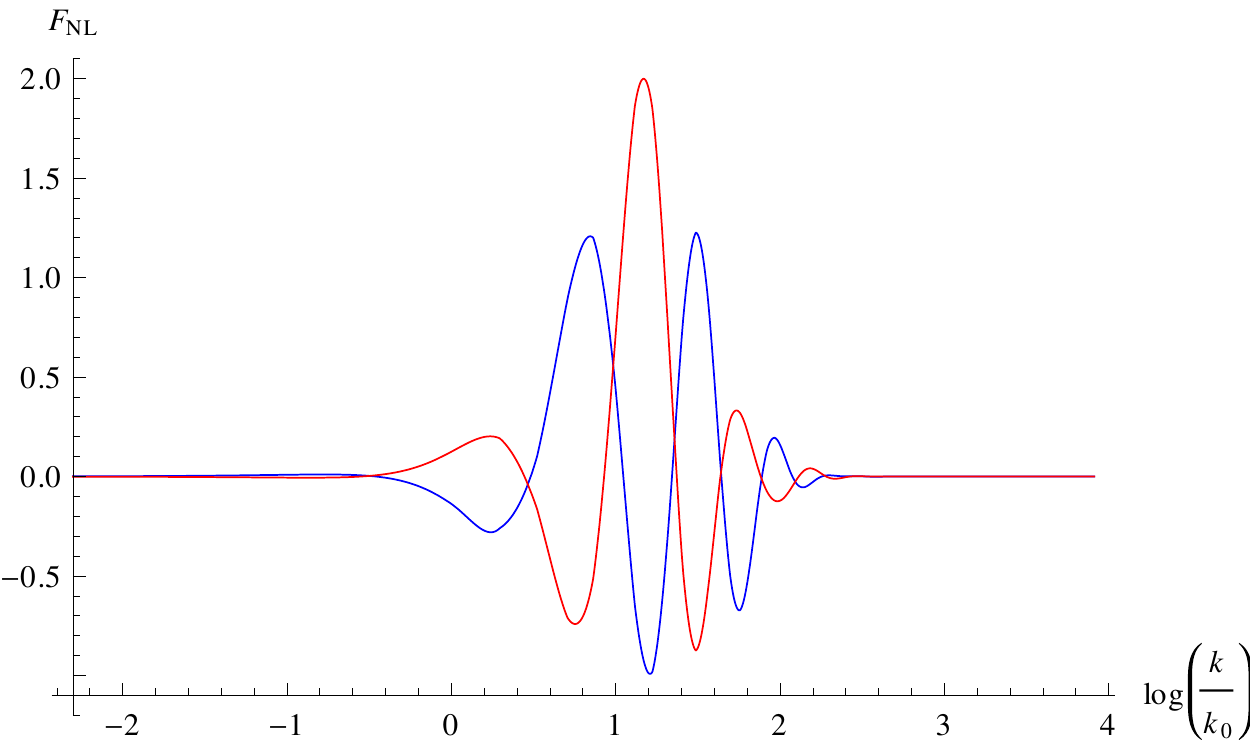}
 \caption{The numerically computed equilateral shape bispectrum is plotted for $\lambda=10^{-11}$ (blue) and $-\lambda=10^{-11}$ (red), $\sigma=0.05$ and $n=1$.}
\label{pmlambdaequilateral}
\end{figure}

\begin{figure}
  \includegraphics[scale=1]{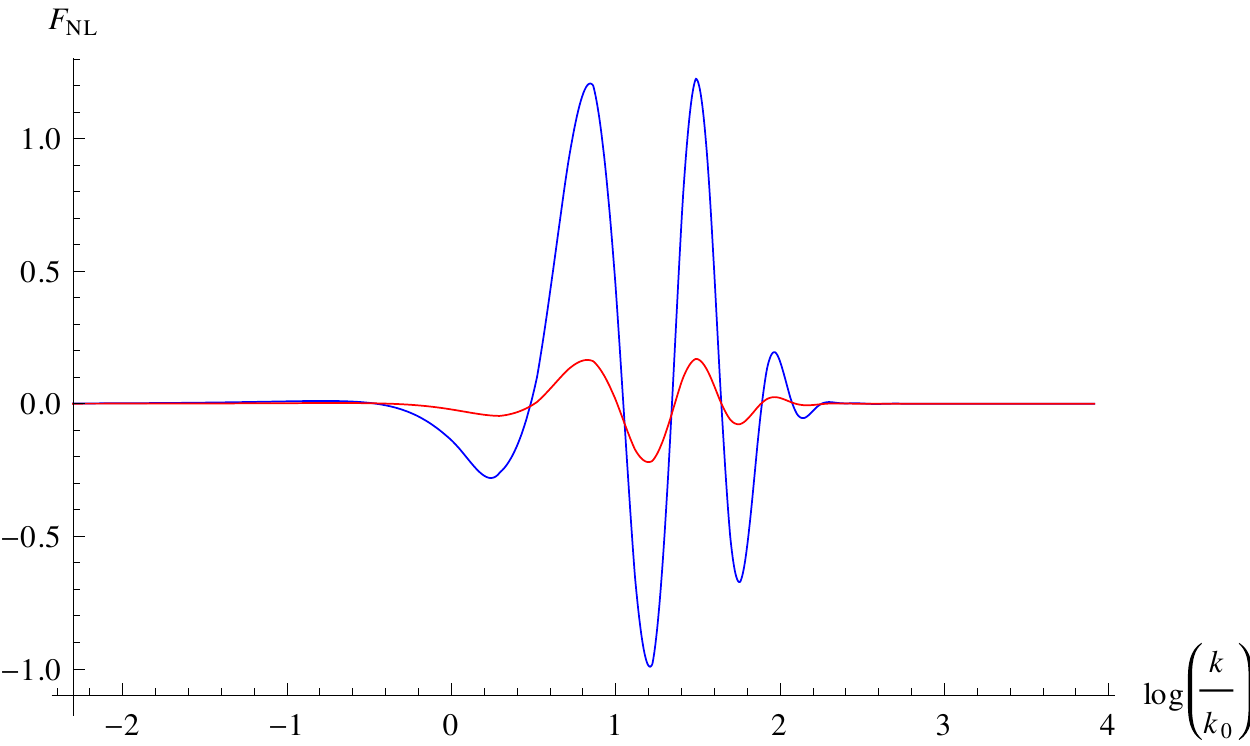}
 \caption{The numerically computed equilateral shape bispectrum is plotted for $\lambda=10^{-11}$ (blue) and $\lambda=10^{-12}$ (red), $\sigma=0.05$ and $n=1$.}
\label{plambdaequilateral}
\end{figure}

We consider inflationary models with a single scalar field and a standard kinetic term with action \cite{linf, inf}
\begin{equation}\label{action1}
  S = \int d^4x \sqrt{-g} \left[ \frac{1}{2} M^2_{Pl} R  - \frac{1}{2}  g^{\mu \nu} \partial_\mu \phi \partial_\nu \phi -V(\phi)
\right],
\end{equation}
where $ M_{Pl} = (8\pi G)^{-1/2}$ is the reduced Planck mass. The variation of the action with respect to the metric  
and the scalar field gives the Friedmann equation and the equation of motion of the inflaton
\begin{equation}\label{ema}
  H^2 \equiv \left(\frac{\dot a}{a}\right)^2= \frac{1}{3 M^2_{Pl}}\left( \frac{1}{2} \dot \phi^2 + V(\phi) \right),
\end{equation}
\begin{equation}\label{emphi}
  \ddot \phi + 3H\dot \phi + \partial_{\phi}V = 0,
\end{equation}
where $H$ is the Hubble parameter and we denote with dots and $\partial_{\phi}$ the derivatives with respect to time and scalar field, respectively. 
The definitions we use for the slow-roll parameters are
\bea \label{slowroll}
  \epsilon \equiv -\frac{\dot H}{H^2} \,\,\,\, , \,\,\,\, \eta \equiv \frac{\dot \epsilon}{\epsilon H}.
\eea
\section{Local features versus branch features}
\label{m}
We consider a single scalar field inflationary model with potential \cite{GallegoThesis}
\bea\label{pot}
V(\phi)&=& V_{0}(\phi) + V_F(\phi) \, , \\
V_F(\phi)&=&\lambda e^{-( \frac{\phi-\phi_0}{\sigma})^{2n}} \, ;\,  n>0 \, , \label{LF}
\eea
where $V_{0}$ is the featureless potential, and we call this type of modification of the slow-roll potential local features (LF).
Many of the features previously studied belong to the category of branch features, that differ from LF because their definition involves step 
functions or their smoothed version, which effectively divide the potential in separate branches.

Some examples of BF are given by the Starobinsky model \cite{Starobinsky:1998mj} 
or its generalizations \cite{Romano:2014kla,whipped} 
\bea
V_F(\phi)&=&\lambda(\phi-\phi_0)^n\theta(\phi-\phi_0)
\eea
The Starobinsky model corresponds to the case $n=1$, for which an analytical solution was first found in \cite{Starobinsky:1998mj,starobinsky}, 
and more recently these studies were followed by \cite{Bousso:2013uia,wiggly}.
The type of features we study in this paper differ from \cite{Romano:2014kla} in the fact that here we study only local features, which only affect 
the potential in a limited range of the field values, while in \cite{Romano:2014kla} the features are not local and modify the potential for an entire 
branch, due to the absence of any dumping factor. 

Some other smoothed versions involving  hyperbolic trigonometric function instead of the Heaviside function have also been studied \cite{Adams}.  They 
can schematically be expressed as
\bea
V_F(\phi)=\lambda \tanh(\frac{\phi-\phi_0}{\Delta}) \,.
\eea
For these models the potential is not only modified around the feature, but for any value in the branch defined by the feature. The direct consequence is that the effects of BF on perturbation modes are not only visible around the scale $k_0$ leaving the horizon around the time of the feature, defined as $\phi(\tau_0)=\phi_0$, but for any scale leaving the horizon when $\phi$ has a value within the feature branch.
This is evident for example from the fact that the spectrum of BF shows a step around $k_0$ \cite{aer}, i.e. the affected scales are all the ones larger (or smaller, depending on the feature) than $k_0$.
On the contrary LF are only affecting the perturbation modes which leave the horizon around $\tau_0$, and consequently the spectrum does not show a step, but a local dumped oscillation and it approaches the featureless spectrum for sufficiently smaller and larger scales.
This is very important because it could allow to model local features of the observed spectrum without affecting other scales.
The different effects of LF and BF on the power spectrum are shown in fig.(\ref{Pplotall}), where it can be seen that they both produce oscillations qualitatively similar around $k_0$, but in the case of BF there is also a step between large and small scales, which is absent for LF.
For BF if the branches of the feature definition were inverted the role of small and large scales would also be accordingly inverted.

In this paper we will consider the case of power law inflation (PLI) to model the featureless behavior
\bea
V_{0}(\phi)=A e^{-\sqrt{\frac{2}{q}}\frac{\phi}{M_{Pl}}}\, .
\eea
While PLI is not in good agreement with CMB data due to high value of the predicted tensor-to-scalar ratio $r$, it can be used as good toy model to show qualitatively the general type of effects produced by LF.
Future works may be devoted to test different potentials $V_0(\phi)$, for direct comparison with data.
\section{Local versus branch features effects}
\begin{figure}
 \begin{minipage}{.45\textwidth}
  \includegraphics[scale=0.6]{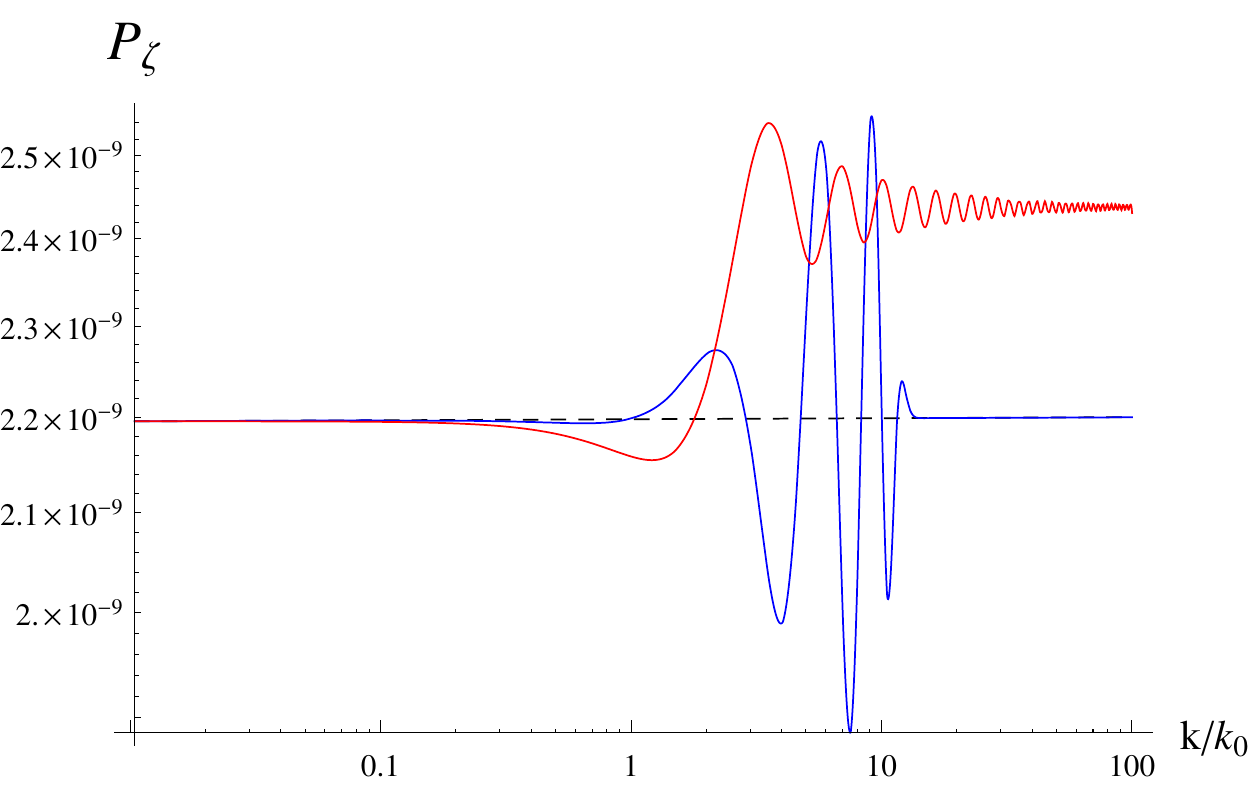}
  \end{minipage}
 \begin{minipage}{.45\textwidth}
  \includegraphics[scale=0.6]{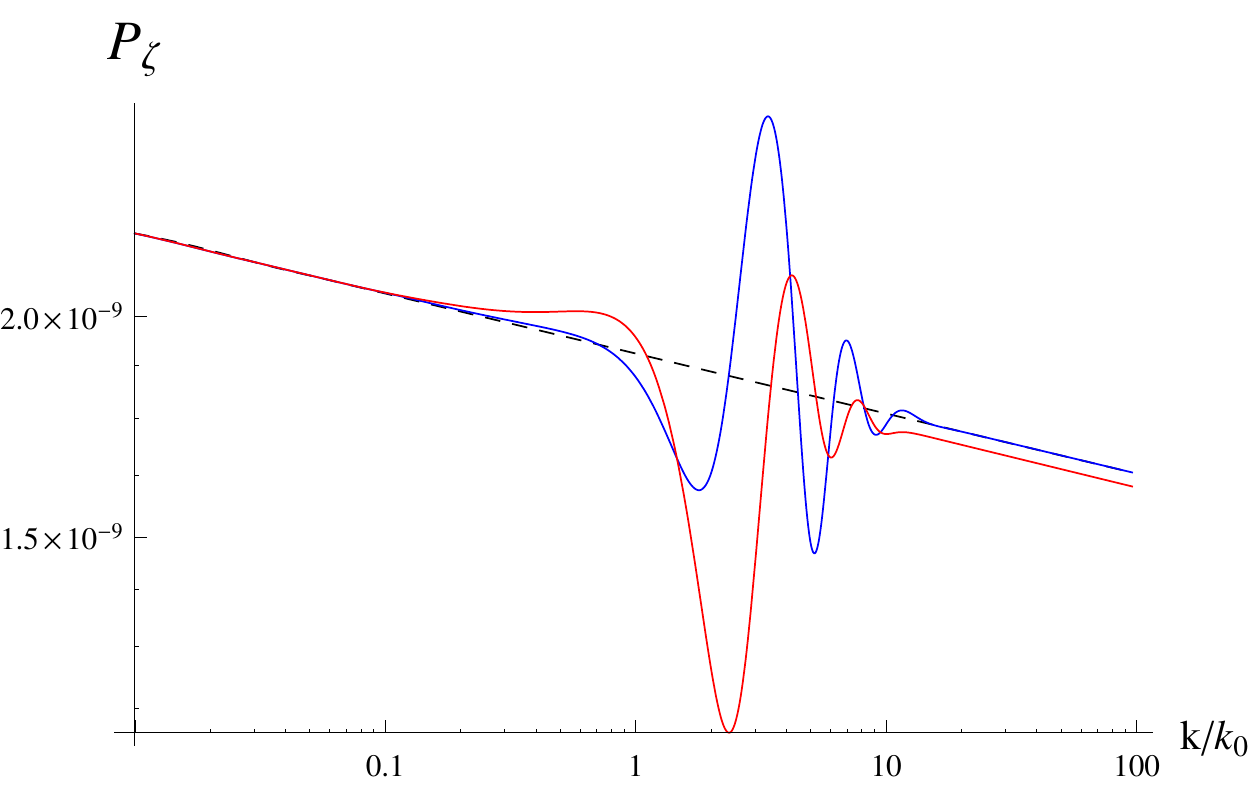}
 \end{minipage}
 \caption{The numerically computed power spectra are plotted as a function of $k/k_0$ for  different potentials. The dashed  lines are the spectra for the featureless potential $V_0(\phi)$, the  blue lines are for the spectrum corresponding to a LF, and the red lines to a BF. 
On the left we have $V_0(\phi)=V_{vac}+\frac{1}{2}m^2 \phi^2$, where $V_{vac}=3.3\times 10^{-13} M_{pl}^3$ and $m=6 \times 10^{-9}M_{pl}$. For LF the potential is $V(\phi)=V_0(\phi) + \lambda e^{-(\frac{\phi-\phi_0}{\sigma})^2}$, with $\lambda=6.5\times 10^{-21}$, $\sigma=10^{-4}$, while the BF is of the type studied before in \cite{Romano:2014kla} corresponding to $V(\phi)=V_0(\phi) + \lambda \theta(\phi_0-\phi)$, with $\lambda=-10 ^{-16}$. \\
On the right we have $V_{0}(\phi)=A e^{-\sqrt{\frac{2}{q}}\frac{\phi}{M_{Pl}}}$, the power law potential. For LF the potential is $V(\phi)=V_0(\phi) + \lambda e^{-(\frac{\phi-\phi_0}{\sigma})^2}$, with $\lambda=-10^{-11}$ and $\sigma=0.05$, while the BF corresponds to  $V(\phi)=V_0(\phi) + \frac{\lambda}{2} \left[1+ \tanh(\frac{\phi_0-\phi}{\sigma})\right]$, with $\lambda=-3\times 10 ^{-11}$ and $\sigma=0.05$. 
The parameters have been chosen so that the different types of features produce oscillations of similar size, so that the effects can be compared consistently.
As it can be seen, depending on the featureless potential $V_0$, there are cases in which the step  between the large and small scale spectrum produced by BF can be important. The oscillation  patterns can also be different, since a single BF has only one phase of slow-roll violation while in a single LF there are two slow-roll violation phases, corresponding to the increasing and decreasing part of the feature.
}
\label{Pplotall}
\end{figure}

One important difference between the effects of BF and LF as mentioned earlier is that BF introduce a step between large and small scales in the spectrum, whose size depends on the featureless model and on the choice of the parameters, in particular $\lambda$. 
The effects of LF and BF producing oscillations of comparable size are shown for different potentials in fig.(\ref{Pplotall}).  As it can be seen, depending on the featureless potential $V_0(\phi)$,  there can be steps of different size in the spectrum. If $\lambda$ is very small the effects of BF and LF could  be phenomenologically indistinguishable, but only a detailed and systematic data analysis can allow to establish  which one of the two categories is observationally preferred, or if they are both compatible with observations.

\begin{figure}
  \includegraphics[scale=1]{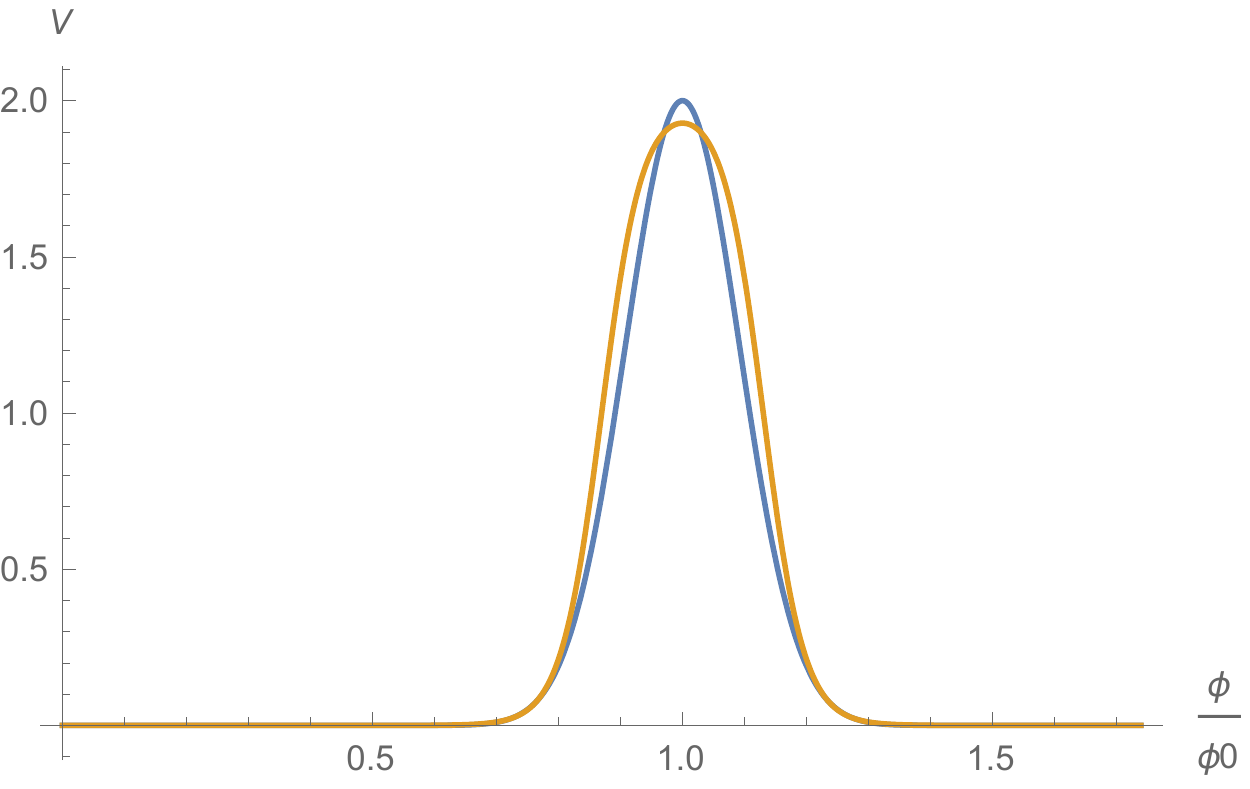}
 \caption{The local feature potential $V_{LF}$ in eq.(\ref{VLF}) and its approximation as the sum of two BF potentials $V_{BF}^1(\phi)+V_{BF}^2(\phi)$ defined in eq.(\ref{VBF12}) are  plotted respectively in blue and in light brown around $\phi_0$ for $\phi_1=1$, $\phi_2=1.2$, $\lambda_1=1$ and $n=1$. This is an example of the fact that appropriate combinations of two BF  can produce modifications of the potential similar to a local feature.}
\label{VLFBF12}
\end{figure}

Another important difference is that LF produces \textit{two slow-roll violation phases}, associated to the increasing and decreasing part of the potential feature, while BF have only one. In this sense a LF can be thought  as the superposition of two BF: for example two smoothed steps with different transition points $\phi_0$ and inverted branches are equivalent to one local smoothed step. The distance between the two transition points is related to the LF $\sigma$ parameter, but each BF has also his own $\sigma$ parameter controlling the smoothness of each BF transition, so the equivalence is not complete, and there can still be differences between the effects of a LF and appropriate combinations of BF, as shown in fig.(\ref{VLFBF12}). 
Schematically, based on general symmetry arguments, we can write
\bea
V_{LF}(\phi)& \approx &V_{BF}^1(\phi)+V_{BF}^2(\phi) \,.
\eea
For example we can approximate a LF as
\bea
V_{LF}(\phi)&=&\lambda e^{-( \frac{\phi-\phi_0}{\sigma})^{2n}} \approx V_{BF}^1(\phi)+V_{BF}^2(\phi) \, \label{VLF}\\
V_{BF}^1(\phi)=\lambda_1 \tanh{\left(\frac{\phi-\phi_1}{\sigma_1}\right)} &\,,& V_{BF}^2(\phi)=\lambda_1 \tanh{\left(\frac{\phi_2-\phi}{\sigma_1}\right)} \label{VBF12}\\
 \phi_0=\frac{\phi_2+\phi_1}{2} \,, \lambda=2 \lambda_1 &\,,& \sigma=\frac{\phi_2-\phi_1}{2} \,, \sigma_1=\frac{\sigma}{2}
\eea
Only a systematic analysis of observational data can determine which is the phenomenologically preferred type of feature, but in general a single BF is expected to produced different effects than a single LF, since this is approximately equivalent to the combination of two single BF.

\section{Curvature perturbations}
From now on we adopt a system of units in which $c=\hbar=M_{Pl}=1$.
The study of curvature perturbations is attained by expanding perturbatively the action with respect to the background $FRLW$ solution \cite{m,hael}. In the comoving gauge the second and third order actions for scalar perturbations are
\bea
\label{s2}
 S_2&=& \int dt d^3x\left[a^3 \epsilon \dot\zeta^2-a\epsilon(\partial \zeta)^2 \right],
\eea
\bea\label{s3}
 S_3=  \int dt d^3x\left[a^3 \epsilon^2 \zeta \dot\zeta^2+ a\epsilon^2 \zeta(\partial \zeta)^2- 2a\epsilon \dot \zeta (\partial\zeta)(\partial \chi)+
 \frac{a^3\epsilon}{2}\dot \eta \zeta^2\dot \zeta \right. \\ \nonumber
 \left. + \frac{\epsilon}{2a}(\partial \zeta)(\partial \chi)\partial^2\chi + \frac{\epsilon}{4a}(\partial^2 \zeta) (\partial \chi)^2+ f(\zeta) \frac{\delta L}{\delta \zeta}\bigg|_1 \right],
\eea
where
\begin{eqnarray}
  \frac{\delta L}{\delta \zeta}\bigg|_1&=& 2a\left(\frac{d\partial^2\chi}{dt}+H \partial^2\chi-\epsilon\partial^2\zeta \right),\\
 f(\zeta) &=& \frac{\eta}{4}\zeta + \mbox{ terms with derivatives on } \zeta,
\end{eqnarray}
and we denote with $\delta L/\delta \zeta|_1$  the variation of the quadratic action with respect to $\zeta$ \cite{m}. 
The Lagrange equations for the second order action give
\begin{equation}
 \frac{\partial}{\partial t}\left(a^3\epsilon \frac{\partial \zeta}{\partial t}\right)- a\epsilon\delta^{ij} \frac{\partial^2 \zeta}{\partial x^i\partial x^j}=0.
\end{equation}
Taking the Fourier transform and using the conformal time $d\tau \equiv dt/a$ we obtain
\begin{equation}\label{cpe}
  \zeta''_k + 2 \frac{z'}{z} \zeta'_k + k^2 \zeta_k = 0,
\end{equation}
where $z\equiv a\sqrt{2 \epsilon}$, $k$ is the comoving wave number, and primes denote derivatives with respect to the conformal time.
    
A similar approach can be adopted for the perturbations of the tensor modes $h_k$, which satisfy the equation
\bea
h''_k+2\frac{a'}{a} h'_k+k^2 h_k=0 \, .
\eea
The power spectrum of scalar perturbations is the Fourier transform of the two-point correlation function of $\zeta$.
For the power spectrum of scalar perturbations we adopt the definition 
\bea
P_{\zeta}(k) \equiv \frac{2k^3}{(2\pi)^2}|\zeta_k|^2,
\eea
and for the power spectrum of tensor perturbations  
\bea
P_{h}(k) \equiv \frac{2k^3}{\pi^2}|h_k|^2\,.
\eea
The tensor-to-scalar ratio is defined as the ratio between the spectrum of tensor and scalar perturbations
\bea
r\equiv \frac{P_h}{P_{\zeta}} \, .
\eea
\section{Calculation of the bispectrum of curvature perturbation}
The bispectrum $B_{\zeta}$ is defined as the Fourier transform of the three-point correlation function as
\begin{equation}
 \Braket{ \zeta(\vec{k}_1, t) \zeta(\vec{k}_2, t)  \zeta(\vec{k}_3, t) }= (2\pi)^3 B_{\zeta}(k_1,k_2,k_3) \delta^{(3)}(\vec{k}_1+\vec{k}_2,\vec{k}_3),
\end{equation}
After a field redefinition, we can re-write the third order action as
\begin{equation}
 S_3=   \int dt d^3x\left[-a^3 \epsilon \eta \zeta \dot\zeta^2-\frac{1}{2}a \epsilon\eta \zeta \partial^2 \zeta \right]\,,
\end{equation}
from which the interaction Hamiltonian can be written in terms of the conformal time as	
\bea
H_{int}(\tau)=  \int d^3x \, \epsilon \eta a \left[ \zeta \zeta'{}^2 + \frac{1}{2} \zeta^2 \partial^2 \zeta \right].
\eea
Finally, the three-point correlation function is given by \cite{m,xc}
\bea
\Braket{\Omega| \zeta(\tau_e,\vec{k}_1) \zeta(\tau_e,\vec{k}_2)  \zeta(\tau_e,\vec{k}_3)|\Omega }= -\mathrm{i} \int_{-\infty}^{\tau_e} \Braket{0|\left[ \zeta(\tau_e,\vec{k}_1) \zeta(\tau_e,\vec{k}_2)  \zeta(\tau_e,\vec{k}_3), H_{int}\right]|0 }\,.
\eea
After substitution we get
\bea \label{b}
  B_{\zeta}(k_1,k_2,k_3)
  = 2 \Im\Bigl[ \zeta(\tau_e,k_1) \zeta(\tau_e,k_2) \zeta(\tau_e,k_3)
  \int^{\tau_e}_{\tau_i}\; d\tau \eta \epsilon a^2 
 \biggl( 2\zeta^*(\tau,k_1) \zeta'{}^*(\tau,k_2)\zeta'{}^*(\tau,k_3)  \\ \nonumber 
 - k^2_1 \zeta^*(\tau,k_1)\zeta^*(\tau,k_2)\zeta^*(\tau,k_3) \biggr)   
+ \mbox{ two permutations of } k_1, k_2,\mbox{ and } k_3  \Bigr]\, ,
\eea
where $\Im$ is the imaginary part.
The integral is computed from $\tau_i$ to $\tau_e$, where $\tau_i$ is some time before $\tau_0$  when the feature effects on the background and perturbations evolution start to be important, and  $\tau_e$ is some time after the horizon crossing, when the modes have frozen \cite{a1,a2,a3, bingo,numerical2013}. 

A commonly used quantity introduced to study non-Gaussianity  is the parameter $f_{NL}$  
\bea\label{fNL}
\frac{6}{5} f_{NL}(k_1,k_2,k_3)\equiv \frac{B_{\zeta}}{\mathbf{P}_{\zeta}(k_1)\mathbf{P}_{\zeta}(k_2)+\mathbf{P}_{\zeta}(k_1)\mathbf{P}_{\zeta}(k_3)+\mathbf{P}_{\zeta}(k_2)\mathbf{P}_{\zeta}(k_3)} \, ,
\eea
where
\begin{equation}
\mathbf{P}_{\zeta} \equiv \frac{2\pi^2}{k^3} P_{\zeta} \, .
\end{equation}
After replacing $\mathbf{P}_{\zeta}$ in eq.(\ref{fNL}) we can  obtain  $f_{NL}$ in terms of our definition of the spectrum $P_{\zeta}(k)$
\bea
f_{NL}(k_1,k_2,k_3)= \frac{10}{3}\frac{(k_1 k_2 k_3)^3}{(2\pi)^4} \frac{B_{\zeta}}{P_{\zeta}(k_1)P_{\zeta}(k_2)k^3_3+P_{\zeta}(k_1)P_{\zeta}(k_3)k^3_2+P_{\zeta}(k_2)P_{\zeta}(k_3)k^3_1} \, .
\eea
In this paper we will use a different quantity to study  non-Gaussianity 
\begin{equation}\label{FNL}
  F_{NL}(k_1,k_2,k_3;k_*)\equiv \frac{10}{3(2\pi)^4}\frac{(k_1 k_2 k_3)^3}{k_1^3+k_2^3+k_3^3}\frac{B_{\zeta}(k_1,k_2,k_3)}{P_{\zeta}^2(k_*)} \, ,
\end{equation}
where $k_*$ is the pivot scale at which the power spectrum is normalized, i.e. $P_{\zeta}(k_*)\approx 2.2\times 10^{-9}$.
When the spectrum is approximately scale invariant our definition of $F_{NL}$ reduces to $f_{NL}$ in the equilateral limit, but in general $f_{NL}$ and $F_{NL}$ are not the same. For example in the squeezed limit they are different, but $F_{NL}$ still provides useful information about the non-Gaussian behavior of $B_{\zeta}$.

\section{Effects of the parameter $n$ }
\subsection{Background}
The parameter $n$ is related to the dumping of the feature, and larger values are associated to a steeper change of the potential, as shown in fig.(\ref{nback}).
The slow-roll parameters show an oscillation around the feature time $\tau_0$ with a larger amplitude for larger $n$, since a steeper potential change is also associated  to larger derivatives of the Hubble parameter as shown in fig.(\ref{nback}).
To better understand the effects on the slow-roll parameter we define the quantity
\bea
\Delta H=H_F-H_0 \,,\\
\eea
where $H_F$ is the Hubble parameter for the model with a feature, and $H_0$ is for the featureless model.
From the definition in eq.(\ref{slowroll}) we can easily see that at leading order in $\Delta H$ we have
\bea
\epsilon_F=\epsilon_0+\Delta \epsilon\,, \\
\Delta\epsilon\approx-\frac{\partial_t\Delta H}{H_0^2}\,,
\eea
where we have defined
\bea
\epsilon_F&=&-\frac{\dot{H_F}}{H_F^2}\,, \\
\epsilon_0&=&-\frac{\dot{H_0}}{H_0^2} \,.\\
\eea
The temporary violation of slow-roll conditions comes from the time derivative of $\Delta H$, so even small changes in the expansion history of the Universe can produce important non-Gaussianities if they happen sufficiently fast. 
In the limit of very large $n$ the feature of the potential tends to a local bump characterized by a very steep transition.
\subsection{Perturbations}
As shown in fig.(\ref{npert}) the tensor-to-scalar ratio $r$ and the spectrum of primordial curvature perturbations show oscillations around the scale $k_0=-1/\tau_0$ with an amplitude which increases for larger $n$. We can understand this from the behavior of $\Delta H$, which has a larger time derivative for larger $n$, because the transition for the potential is also steeper.
As seen in fig.(\ref{nequilateral}) the equilateral limit of the bispectrum also shows oscillations around $k_0$, which are larger for larger $n$, for the same reason given above.
It is important to observe that both the spectrum and the bispectrum are only affected in a limited range of scales, since this is a LF.
For BF, instead, the change affects all the scales before of after the feature \cite{aer,Romano:2014kla}, because the potential is modified in the entire branch.

\section{Effects of the parameter $\sigma$ }
\subsection{Background}
The parameter $\sigma$ determines the size of the range of field  values where the potential is affected by the feature, as shown in fig.(\ref{sigmaback}). The slow-roll parameters are smaller for larger $\sigma$ since a larger width of the feature tend to reduce the time derivative of the Hubble parameter.
In this case, in fact, the modification of the potential is also associated to smaller derivatives with respect to the field, 
since the shape of the potential is less steep.
\subsection{Perturbations}
As shown in fig.(\ref{sigmapert})  the spectrum of primordial curvature perturbations and the tensor-to-scalar ratio $r$ have oscillations around $k_0$, whose amplitude is larger for smaller $\sigma$, because in this case the potential changes faster and consequently the slow-roll parameters are larger.
In fig.(\ref{sigmaequilateral}) we can see that the equilateral limit of the bispectrum also presents oscillations around $k_0$ with larger amplitude for smaller $\sigma$.
Both for the spectrum and bispectrum the effects are confined in a limited range of the scales, differently from what we would have with BF.

\section{Effects of the parameter $\lambda$ }
\subsection{Background}
The parameter $\lambda$ controls the magnitude  of the potential modification, as shown in figs.(\ref{pmlambdaback}-\ref{plambdaback}). For larger absolute values of $\lambda$, the slow-roll parameters are larger in absolute value, since a larger feature of the potential induces a larger time derivative of the Hubble parameter. 
The sign of $\lambda$ produce opposite and symmetric effects, since it  implies an opposite sign for the derivative of the potential with respect to the field, and consequently of the Hubble parameter with respect to time.

\subsection{Perturbations}
As shown in figs.(\ref{pmlambdapert},\ref{plambdapert},\ref{pmlambdaequilateral},\ref{plambdaequilateral}), larger absolute values of $\lambda$ produce oscillations with a larger amplitude for the tensor-to-scalar ratio $r$, the spectrum and bispectrum around $k_0$.
Features with the same absolute value and opposite sign of $\lambda$ correspond to oscillations that are symmetric with respect to the featureless spectrum and bispectrum.

\section{Effects on the CMB temperature and polarization spectrum}

In this section we present the effects of the feature on the CMB temperature and polarization spectrum.
To study how the feature impacts on the CMB spectra we modified the Boltzmann equations solver
CAMB (Code for Anisotropies in the Microwave Background) \cite{Lewis:1999bs}
to use the modified primordial power spectrum instead of the usual power-law expression
$P_s(k)=A_s\, (k/0.05\,\textrm{Mpc}^{-1})^{n_s-1}$,
where $A_s$ is the normalization and $n_s$ is the tilt of the power-law spectrum.

The CMB spectrum is the convolution of the power spectrum of initial perturbations with the transfer function,
which is calculated by CAMB assuming the standard cosmological model.
We fix all the cosmological parameters to the Planck 2015 best-fit values \cite{Ade:2015xua}.

\begin{table}
\begin{tabular}{l||c|c|c}
Label & $\lambda$ & $\sigma$ &$n$\\
\hline
Feature A	&    $ 10^{-11}$ & 0.05 & 1\\
Feature B	&    $ 10^{-11}$ & 0.05 & 2\\
Feature C	&    $-10^{-11}$ & 0.05 & 1\\
Feature D	&    $ 10^{-11}$ & 0.1  & 1\\
Feature E   &    $ 10^{-12}$ & 0.05 & 1\\
\end{tabular}
\caption{\label{Pint}
The values of $\lambda$, $\sigma$ and $n$ we used for the different spectra in figs.(\ref{clTT}-\ref{clBB}).
}
\end{table}

\begin{figure}
	\centering
  \includegraphics[scale=0.5]{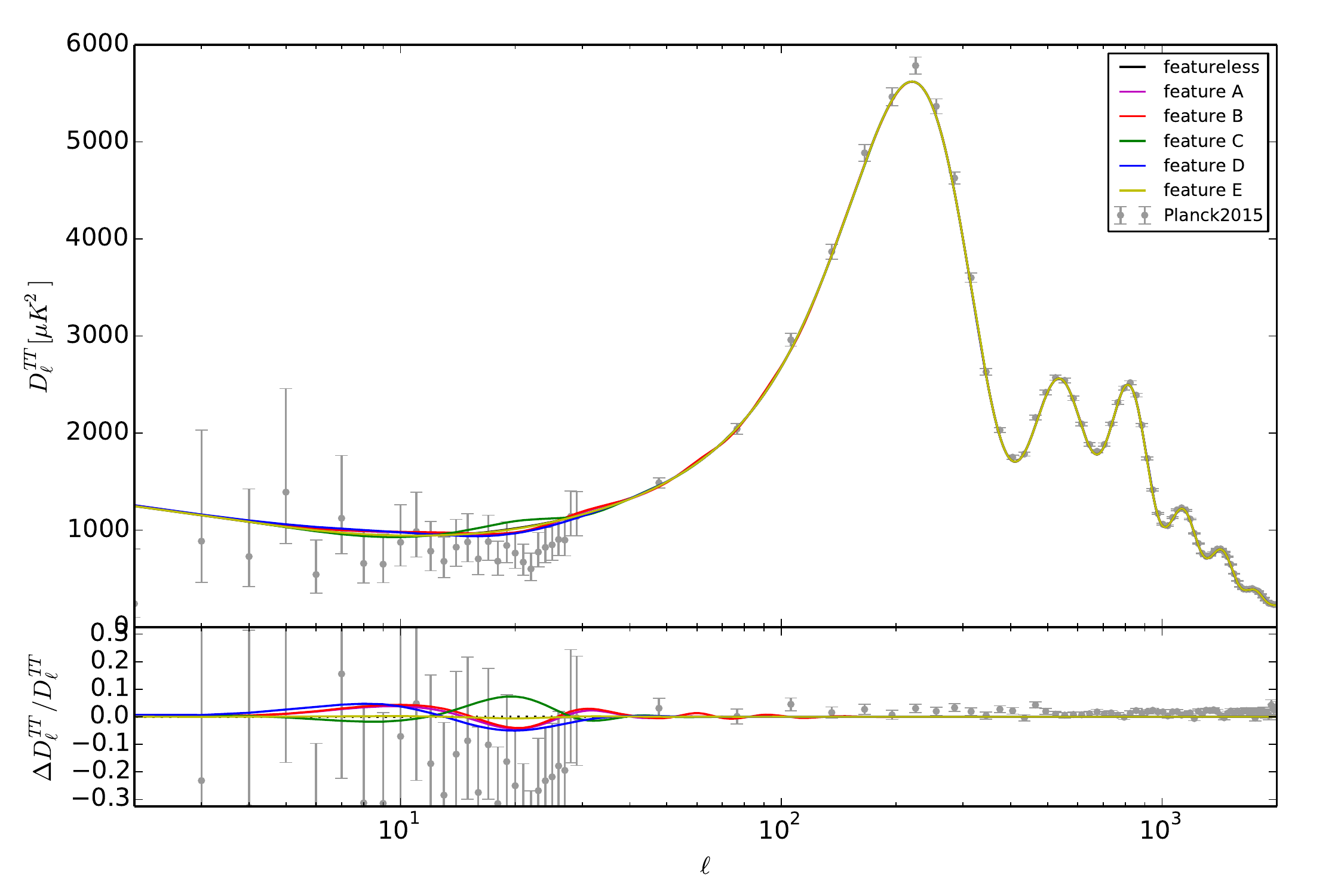}
 \caption{We plot the $D_l^{TT}=\ell(\ell+1) C_\ell^{TT}/(2\pi)$ spectrum in units of $\mu K^2$ with respect to the multipole $l$, and the relative difference with respect to the featureless behavior.
 The solid black lines correspond to the featureless behavior.
 The values of $\lambda$, $\sigma$ and $n$ we used for the different curves are listed in tab.\ref{Pint}.}
\label{clTT}
\end{figure}

\begin{figure}
	\centering
  \includegraphics[scale=0.5]{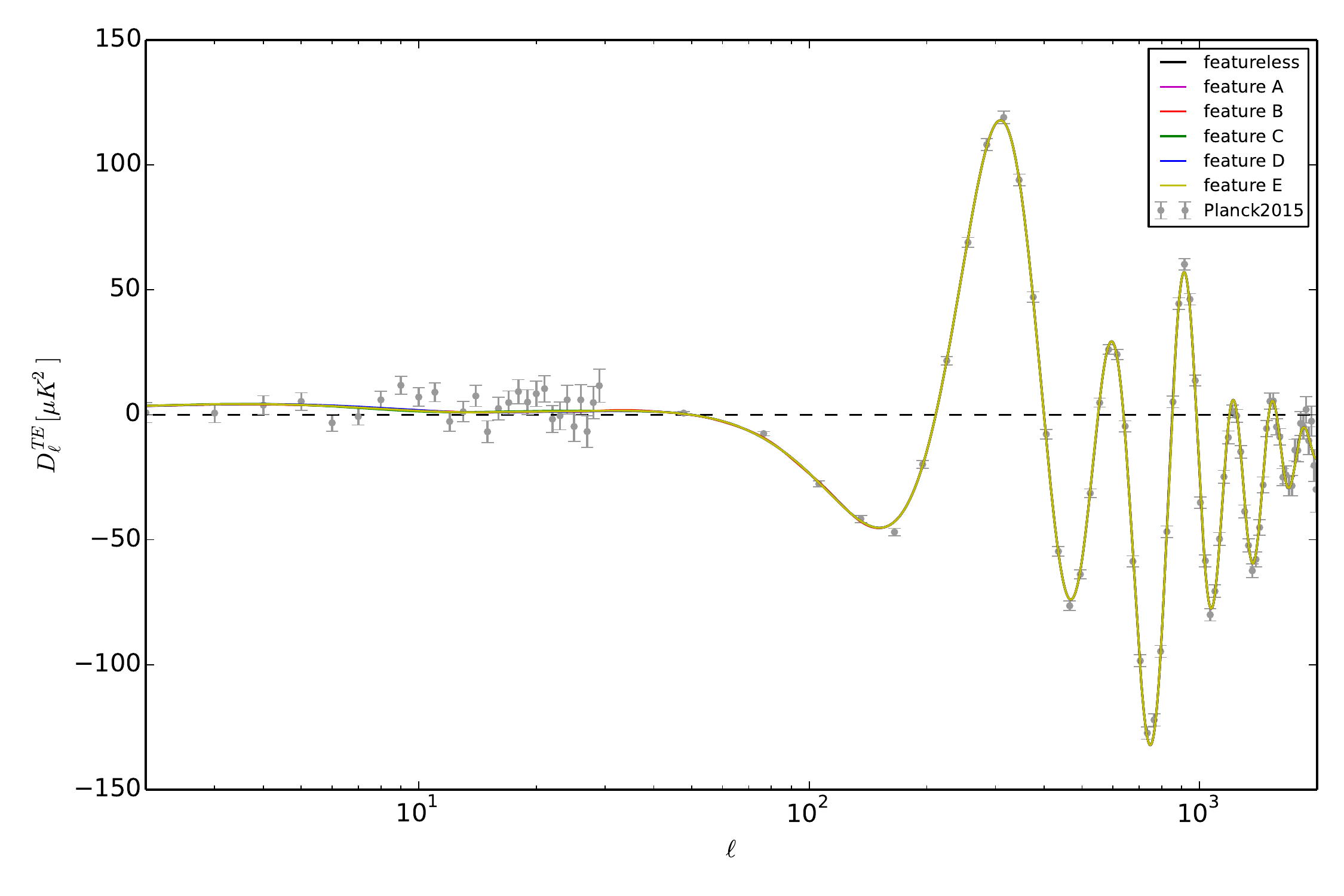}
 \caption{We plot the $D_l^{TE}=\ell(\ell+1) C_\ell^{TE}/(2\pi)$ spectrum in units of $\mu K^2$ with respect to the multipole $l$.
 The solid black lines correspond to the featureless behavior.
 The values of $\lambda$, $\sigma$ and $n$ we used for the different curves are listed in tab.\ref{Pint}.}
\label{clTE}
\end{figure}

\begin{figure}
	\centering
  \includegraphics[scale=0.5]{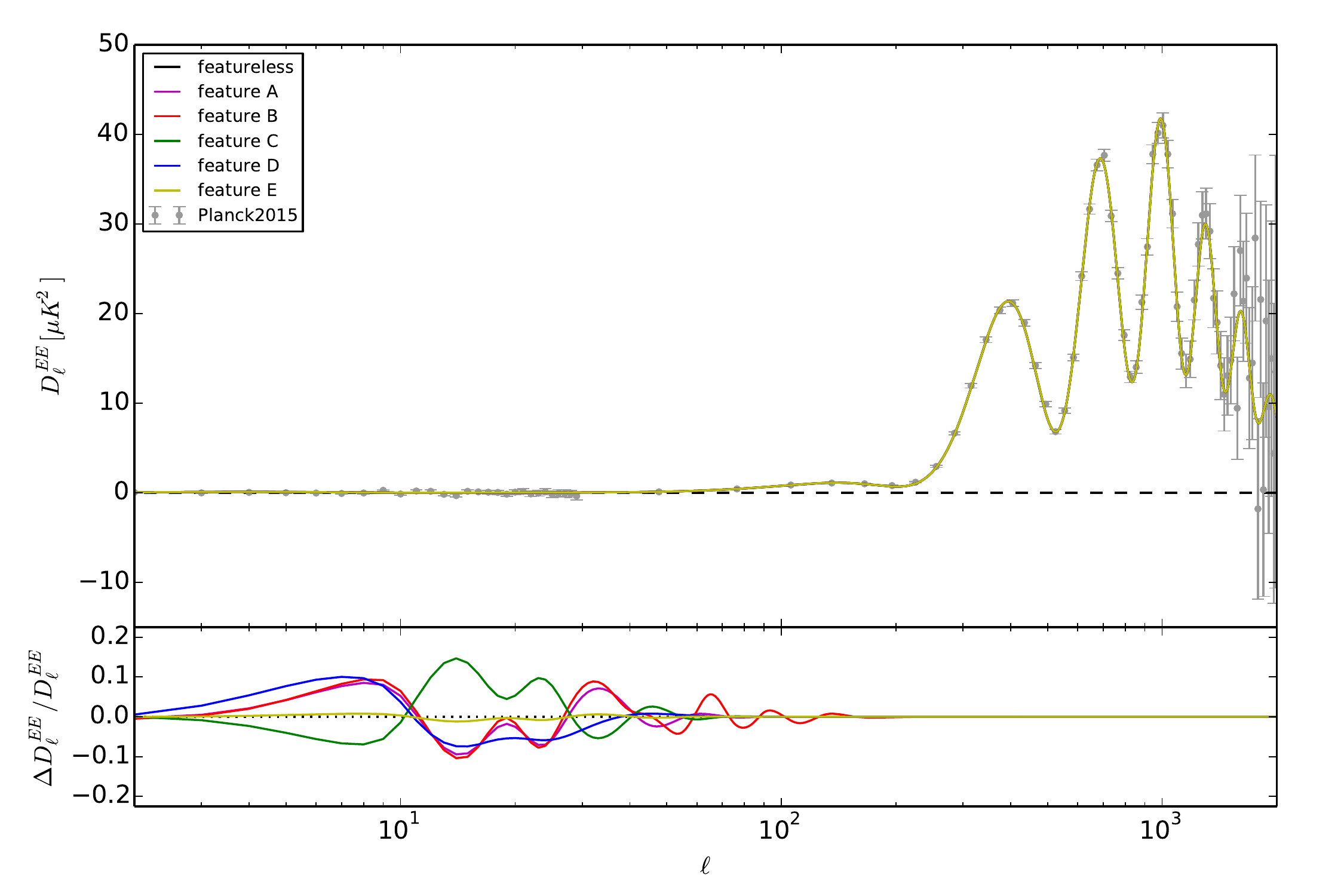}
 \caption{We plot the $D_l^{EE}=\ell(\ell+1) C_\ell^{EE}/(2\pi)$ spectrum in units of $\mu K^2$ with respect to the multipole $l$, and the relative difference with respect to the featureless behavior.
 The solid black lines correspond to the featureless behavior.
 The values of $\lambda$, $\sigma$ and $n$ we used for the different curves are listed in tab.\ref{Pint}.}
\label{clEE}
\end{figure}

\begin{figure}
	\centering
  \includegraphics[scale=0.5]{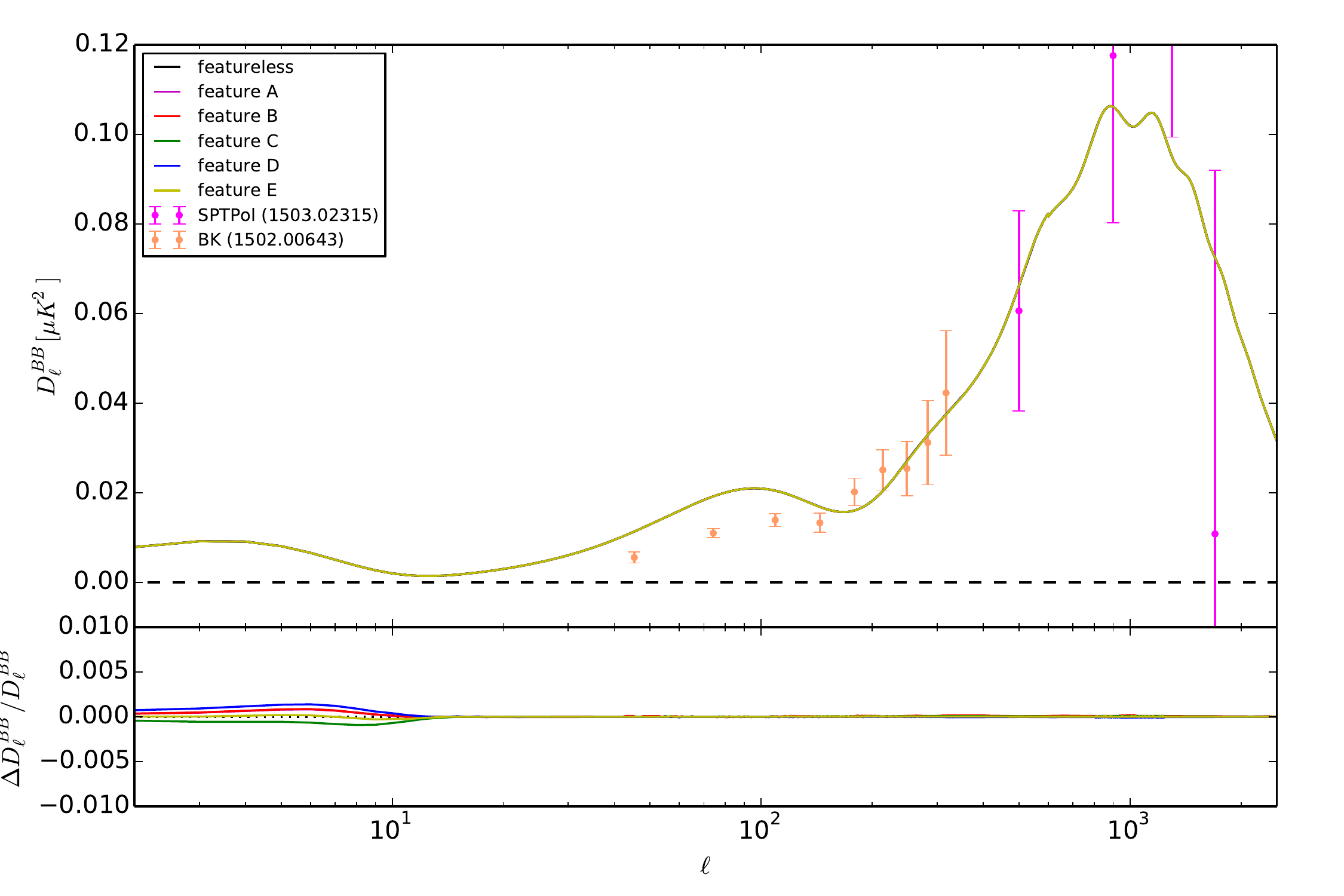}
 \caption{We plot the $D_l^{BB}=\ell(\ell+1) C_\ell^{BB}/(2\pi)$ spectrum in units of $\mu K^2$ with respect to the multipole $l$, and the relative difference with respect to the featureless behavior.
 The solid black lines correspond to the featureless behavior.
 The values of $\lambda$, $\sigma$ and $n$ we used for the different curves are listed in tab.\ref{Pint}.}
\label{clBB}
\end{figure}

In figs.(\ref{clTT}-\ref{clBB}) we show the CMB spectra obtained with different combinations of the parameters
$\lambda$, $\sigma$ and $n$:
in tab.\ref{Pint} we list the values of the feature parameters for each combination.
We show the spectra in terms of the quantity $D_\ell=\ell(\ell+1) C_\ell/(2\pi)$.
For each of the
temperature (TT, fig.(\ref{clTT})),
E-mode polarization (EE,  fig.(\ref{clEE})) and
B-mode polarization (BB,  fig.(\ref{clBB}))  auto-correlation power spectra
we plot the $D_\ell$ spectra (top) and the relative difference with respect to the featureless spectrum (bottom)
for a large range of multipoles.
In  fig.(\ref{clTE}) we show also the TE cross-correlation power spectra.
The spectra are compared with the most recent experimental data:
we plot the Planck data \cite{Adam:2015rua} for the TT, TE and EE spectra
and the points obtained by the
SPTPol experiment \cite{Keisler:2015hfa} and
the Bicep-Keck (BK) collaboration \cite{Ade:2015fwj} data for the BB spectrum.
The SPTPol data provide the first detection of the B-modes generated by the lensing of E-modes perturbations.
We recall that the BK data plotted here contain a significant contamination from B-modes emitted by dust \cite{Ade:2015tva}.

From the various plots it is possible to see how the feature can change the predicted CMB spectra.
In particular, the most significant variations are in the TT and in the EE spectra,
where relative differences of the order of $10-15\%$ are visible,
while the presence of the feature has a very small impact on the B-modes spectrum.
Choosing the values listed in tab.(\ref{Pint}) for the parameters describing the feature,
and
$k_0=5\times 10^{-4}\,\textrm{Mpc}^{-1}$, $A_s=2.2\times 10^{-9}$ and $n_s=0.967$ we can see from fig.(\ref{clTT})
that the effects of some feature can partially reproduce 
the dip at $\ell\simeq20$ in the TT spectrum. 
Further studies involving data fitting can determine more accurately the values of the parameters 
which provide the best explanation of the observed deviation of the power spectrum from the power law form,
but they go beyond the scope of this paper, and will be reported separately.
From figs.(\ref{clTT}-\ref{clBB}) it is also clear that the presence of the LF affects only a part of the CMB spectra, while far from the multipoles corresponding to the feature scale the spectra are equal to those obtained in the featureless case.


\section{Conclusions}
We have studied the effects of local features of the inflation potential on the spectrum and bispectrum of single field inflationary models with a canonical  kinetic term.
These features only modify the potential in a limited range of the scalar field values, and consequently only affect the spectrum and bispectrum in a narrow range of scales, which leave the horizon during the time interval corresponding to the modification of the potential.
This is different from branch type features which effectively divide the potential into separate branches, because they involve a step-like function in their definition.
Some examples of branch type features are the Starobinsky model \cite{Starobinsky:1998mj} and its generalizations \cite{Romano:2014kla, whipped}.
In BF models the spectrum can, for example, exhibit a step, reminiscent of the branch-type potential modifications. For local features there is no step, and the spectrum returns to the featureless form for scales sufficiently larger or smaller than $k_0$.

The tensor-to-scalar ratio $r$, the spectrum and the bispectrum of primordial curvature perturbations are affected by the feature, showing modulated oscillations 
which are dumped for scales larger or smaller than $k_0$.
The amplitude of the oscillations depends on the parameters defining the local feature, and the effects are larger when the potential modification is steeper, since in this case there is a stronger violation of the slow-roll conditions.

We have also computed the effects of the features on the CMB temperature and polarization spectra, showing how an appropriate choice of parameters can produce effects in qualitative agreement with the observational CMB data.
Due to this local type effect these features could be used to model phenomenologically local glitches of the spectrum, without affecting  other scales, and it will be interesting in the future to perform a detailed observational data fitting analysis using the new CMB data of the Planck mission.

\acknowledgments
This work was supported by the European Union (European Social Fund, ESF) and Greek national funds under the “ARISTEIA II” Action.
and the Dedicacion exclusica and Sostenibilidad programs at UDEA, the UDEA CODI
projects IN10219CE and 2015-4044.
The work of S.G. was supported by the Theoretical Astroparticle Physics research Grant No. 2012CPPYP7 under the Program PRIN 2012 funded by the Ministero dell'Istruzione, Universit\`a e della Ricerca (MIUR).

\bibliography{Bibliography}
\bibliographystyle{h-physrev4}
\end{document}